\documentclass[preprint,authoryear,3p]{elsarticle}

\newcommand{\ljump}{\llbracket}
\newcommand{\rjump}{\rrbracket}
\newcommand\bx{\mbox{\boldmath $x$}}
\newcommand\be{\mbox{\boldmath $e$}}
\newcommand\bn{\mbox{\boldmath $n$}}
\newcommand\bxi{\mbox{\boldmath $\xi$}}
\newcommand\bY{\mbox{\boldmath $Y$}}
\newcommand{\bmM}{\mbox{\boldmath $\mathcal{M}$}}

\newcommand{\jump}[1]{\ljump #1 \rjump}
\newcommand{\av}[1]{\langle #1 \rangle}

\usepackage{amsmath}
\usepackage{amssymb}
\usepackage{stmaryrd}
\usepackage{epstopdf}
\usepackage{pstricks}
\usepackage{pst-node}
\definecolor{corr}{RGB}{255,1,1}   %----- Makes the most recent amendments red.
\journal{International Journal of Solids and Structures}

\begin{document}

\begin{frontmatter}



\title{Perturbation analysis for an imperfect interface crack problem using weight function techniques}


\author[aber]{A. Vellender\corref{cor1}}
\author[aber,enginsoft]{G. Mishuris}
\author[trento]{A. Piccolroaz}

\address[aber]{Institute of Mathematics and Physics, Aberystwyth University, Physical Sciences Building, Aberystwyth, Ceredigion, Wales, SY23 3BZ}
\address[trento]{Department of Civil, Environmental and Mechanical Engineering, University of Trento, Via Mesiano 77, 38123 Trento, Italy}
\address[enginsoft]{Enginsoft Trento, Via della Stazione 27 - fraz. Mattarello, 38123, Trento, Italy}
 \cortext[cor1]{Corresponding author. {\em Tel}: +44 (0)1970622776. {\em Email address}: asv2@aber.ac.uk (A. Vellender)}

\begin{abstract}
We analyse a problem of anti-plane shear in a bi-material plane containing a semi-infinite crack situated on a soft imperfect interface. The plane also contains a small thin inclusion (for instance an ellipse with high eccentricity) whose influence on the propagation of the main crack we investigate. 
%The problem can be considered as modelling bimaterial ceramics which are joined with a thin adhesive substance. 
An important element of our approach is the derivation of a new weight function (a special solution to a homogeneous boundary value problem) in the imperfect interface setting. The weight function is derived using Fourier transform and Wiener-Hopf techniques and allows us to obtain an expression for an important constant $\sigma_0^{(0)}$ (which may be used in a fracture criterion) that describes the leading order of tractions near the crack tip for the unperturbed problem. We present computations that demonstrate how $\sigma_0^{(0)}$  varies depending on the extent of interface imperfection and contrast in material stiffness. We then perform perturbation analysis to derive an expression for the change in the leading order of tractions near the tip of the main crack induced by the presence of the small defect, whose sign can be interpreted as the inclusion's presence having an amplifying or shielding effect on the propagation of the main crack.
% This derivation uses Betti's identity to relate physical fields to the weight function; the presence of the imperfect interface introduces new challenges. We present computations that demonstrate how $\sigma_0$, which may be used in a fracture criterion, varies depending on the extent of interface imperfection and the contrast in stiffness between the two matierials. We also draw comparisons against the previously studied analogous perfect interface case. We then move on to consider the perturbed problem. In particular, the Betti identity again allows us to use the weight function to derive an expression for $\Delta \sigma_0$, the change in the leading order of tractions near the tip of the main crack induced by the presence of the small defect. The sign of $\Delta \sigma_0$ can be interpreted as the inclusion's presence having an amplifying or shielding effect on the propagation of the main crack.
\end{abstract}


\end{frontmatter}

% \linenumbers


\section{Introduction}
% In this chapter we consider a problem of anti-plane shear in the whole plane (as opposed to the strip geometries in which we have hitherto formulated problems), with different materials occupying the regions above and below the crack line. The geometry considered contains a semi-infinite crack situated along an imperfect interface; we will formulate and solve a weight funtion problem in this geometry. By using Betti's identity in the imperfect interface case, we will use the weight function to derive important constants in a related physical problem. We then conduct perturbation analysis to determine how the presence of small defects in the material affects the stresses near the main crack tip.
In this paper we present a method to evaluate important constants which describe the behaviour of physical fields near crack tips in a perturbed problem set in a domain containing an imperfect interface. 

Imperfect interfaces account for the fact that the interface between two materials is almost never sharp. \cite{Atkinson} accounted for this observation by by placing a very thin strip of a homogeneous material in the model between two larger bodies with different elastic moduli to that of the strip. If the thin layer is considered to be either much softer or stiffer than the main bodies, its presence can be replaced in models by transmission conditions, whose derivation can be found for example in \cite{Antipov2001} for a soft imperfect interface, or \cite{Mish2006} for a stiff imperfect interface. We shall consider only soft imperfect interfaces in the present paper.

\cite{KlarbringMovchan} presented an asymptotic model of adhesive joints in a layered structure. \cite{Mish2001a} found the asymptotic behaviour of displacements and stresses in a vicinity of the crack tip situated on a soft imperfect interface between two different elastic materials, where the non-ideal interface is replaced by non-ideal transmission conditions. For such a case, the asymptotics are of a markedly different form to the perfect interface case, in which components of stress exhibit a square root singularity at the crack tip; such behaviour is not present for imperfect interface cracks.

A key element of our approach will be the derivation of a new weight function. The concept of weight functions was introduced by \cite{Bueckner1970}. In the perfect interface setting these provide weights for the loads applied to the crack surfaces such that their weighted integrals over the crack surfaces provide the stress intensity factors at a certain point. \cite{Vellender2011a} modified the weight function technique to yield similarly useful asymptotic constants that characterise stress fields near crack tips along an imperfect interface.

A survey of macro-microcrack interaction problems can be found in \cite{Petrova}. Of particular relevance is the recent manuscript of \cite{PiccInclusions} which examines an analogous problem to that presently considered with a perfect interface in place of the imperfect interface. The approach in that paper utilises the dipole matrix approach of \cite{Movchanmovchan} to construct an asymptotic solution that takes into account the presence of a micro-defect such as a small inclusion. The present paper seeks to adapt this approach to the imperfect interface setting.

\section{Structure and summary of main results}
We adopt the following structure for the paper. We first formulate the physical problem before giving the weight function problem formulation. Fourier transform techniques allow us to obtain a Wiener-Hopf type problem for the weight function, whose kernel we factorise in a computationally convenient fashion. The Wiener-Hopf equation is solved to yield expressions for the weight function and comparisons are drawn between the perfect and imperfect interface weight function problems.

We then use the reciprocal theorem (Betti formula) in the spirit of \cite{Willis} to relate the sought physical solution to the weight function. The presence of imperfect interface transmission conditions alters properties of the functions in the Betti identity and so different analysis is required. The application of Betti's identity enables us to find an expression for the leading order of tractions $\sigma_0^{(0)}$ near the crack tip in terms of the new weight function and the imposed arbitrary tractions prescribed on the faces of the crack:
\begin{equation}
 {\sigma_0^{(0)}}=\frac{1}{2}\sqrt{\frac{\mu_0}{\pi}}\int\limits_{-\infty}^\infty\xi\left(\bar{\ljump U\rjump}(\xi)\bar{\langle p\rangle}(\xi)+\bar{\langle U\rangle}(\xi)\bar{\ljump p\rjump}(\xi)\right)\mathrm{d}\xi.
\end{equation}
Here, bars denote Fourier transform, $\mu_0$ is a constant depending on the material parameters and extent of interface imperfection, $\jump{U}$ and $\av{U}$ are respectively the jump and average of the weight function across the crack/interface line, and $\jump{p}$ and $\av{p}$ are the jump and average of the tractions prescribed on the crack faces.

In Section \ref{section:pert}, we perform perturbation analysis to determine the impact on the tractions near the crack tip of the presence of a small inclusion. The asymptotic solution is sought in the form
\begin{equation}
u(\bx,\varepsilon) = u^{(0)}(\bx)  + \varepsilon W^{(1)}(\bxi) + \varepsilon^2 u^{(1)}(\bx) + o(\varepsilon^2), \quad \varepsilon \to 0,
\end{equation}
where $u^{(0)}$ is the unperturbed physical displacement solution (the solution with no inclusion present), $\varepsilon W^{(1)}$ is a boundary layer concentrated near the inclusion and $\varepsilon^2 u^{(1)}$ is introduced to fulfil the original boundary conditions on the crack faces and along the imperfect interface. This enables us to find the first order variation in the crack tip tractions; we expand the constant $\sigma_0$ as
\begin{equation}
\sigma_0 = \sigma_0^{(0)} + \varepsilon^2 \Delta \sigma_0 + o(\varepsilon^2), \quad \varepsilon \to 0,
\end{equation}
and use Betti identity arguments to derive an expression for $\Delta \sigma_0$ (see (\ref{deltaa0expression})). This is interpreted physically as the change in traction near the crack tip induced by the inclusion's presence; as such we say that the sign of $\Delta \sigma_0$ for any given positioning and configuration of the inclusion either shields or amplifies the propagation of the main crack. {Note that for the unpeturbed setup (with no inclusion present) $\sigma_0=\sigma_0^{(0)}$ and so we will naturally drop the superscript when referring to the quantity corresponding to the unperturbed problem.}

We conclude the paper by presenting numerical results in Section \ref{section:numerical}. In particular we show how $\sigma_0^{(0)}$ varies depending on the extent of interface imperfection and choice of material contrast parameter for different loadings. These computations are performed for point loadings that are chosen to be illustrative of the suitability of our method to asymmetric self-balanced loadings. We further propose a method of comparing $\sigma_0^{(0)}$ with stress intensity factors from the analogous perfect interface problem and find agreement as the extent of interface imperfection tends towards zero. We also present computations that show the sign of $\Delta \sigma_0$ for varying location and orientation of the micro-defect.

\section{Formulation of physical and weight function problems}
\subsection{Physical formulation}\label{plane:section:physicalformulation}
We consider an infinite two-phase plane with an imperfect interface positioned along the positive $x$-axis. A semi-infinite crack is placed occupying the line $\{(x,y):x<0,y=0\}$. We refer to the half-planes above and below the crack and interface respectively as $\Pi^{(1)}$ and $\Pi^{(2)}$. The material occupying $\Pi^{(j)}$ has shear modulus $\mu_j$ and mass density $\rho_j$ for $j=1,2$. The anti-plane shear displacement function $u$ satisfies the Laplace equation
\begin{equation}\label{plane:laplace}
 \nabla^2u(x,y)=0.
\end{equation}
\begin{figure}
\begin{center}
      \includegraphics[width=0.45\linewidth]{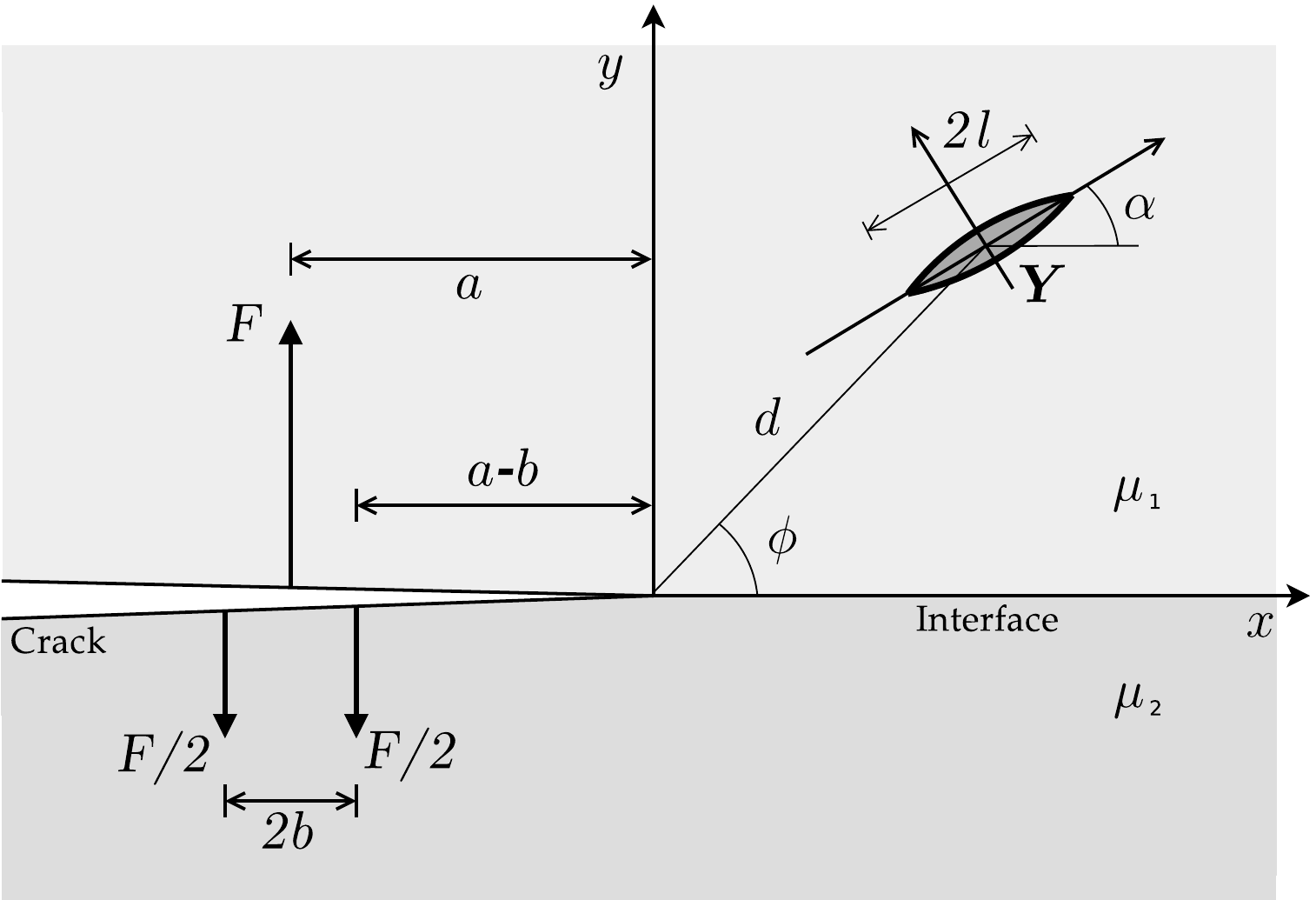}
      \end{center}
\caption{Geometry for the physical setup. The crack tip is placed at the origin of an infinite plane composed of materials with shear modulus $\mu_j$ occupying half-planes $\Pi^{(j)}$ above and below the crack and imperfect interface for $j=1,2.$ The central point $\mathbf{Y}$ of a micro-defect is situated at a distance $d$ from the tip of the main crack.}\label{plane:figure:physicaldefectsetup}
\end{figure}
The plane also contains a micro-defect whose centre is at the point $\bY$; we will consider in particular elliptic inclusions although other types of defect may be incorporated into the model provided a suitable dipole matrix can be obtained (see for example \cite{PiccInclusions} in which micro-cracks and rigid line inclusions are considered). The defect $g_\varepsilon$ has shear modulus $\mu_{\text{in}}$, is placed at a distance $d$ from the crack tip, makes an angle $\phi$ with the imperfect interface and is oriented at an angle $\alpha$ to the horizontal as shown in Figure \ref{plane:figure:physicaldefectsetup}. The value of $\mu_{\text{in}}$ may be greater than or less than the value of $\mu_{\text{out}}$ (which may be $\mu_1$ or $\mu_2$ depending where the defect is placed), and so both stiff and soft defects can be considered.

We assume continuity of tractions across the crack and interface, and introduce imperfect interface conditions ahead of the crack:
\begin{equation}\label{plane:contoftractions}
 \mu_1\left.\frac{\partial u}{\partial y}\right|_{y=0+}=\mu_2\left.\frac{\partial u}{\partial y}\right|_{y=0-},\quad {x>0},
\end{equation}
\begin{equation}
 \ljump u\rjump-\kappa\mu_1\left.\frac{\partial u}{\partial y}\right|_{y=0+}=0,\quad x>0,
\end{equation}
where the notation $\ljump u\rjump$ defines the jump in displacement across $x=0$, i.e. 
\begin{equation}
\ljump u\rjump(x)=u_1(x,0^+)-u_2(x,0^-). 
\end{equation}
The parameter $\kappa>0$ describes the extent of imperfection of the interface, with larger $\kappa$ corresponding to more imperfect interfaces. We further impose prescribed tractions $p_\pm$ on the crack faces:
\begin{equation}
 \mu_1\left.\frac{\partial u}{\partial y}\right|_{y=0+}=p_+(x),\quad \mu_2\left.\frac{\partial u}{\partial y}\right|_{y=0-}=p_-(x); \quad x<0.
\end{equation}
These tractions are assumed to be self-balanced; that is
\begin{equation}
 \int\limits_{-\infty}^0{p_+(x)}\mathrm{d}x-\int\limits_{-\infty}^0{p_-(x)}\mathrm{d}x=0,
\end{equation}
and it is further assumed that $p_\pm(x)$ vanish in a neighbourhood of the crack tip. Although the techniques we will establish can be applied to any permissible loading, we will particularly focus our attention on the case where these loadings are point loadings, with a loading on the upper crack face positioned at $x=-a$ (where $a>0$) balanced by two equal point loadings on the lower crack face positioned at $x=-a-b$ and $x=-a+b$, where $0<b<a$. This loading makes computations more difficult to perform than for the smooth loadings considered by \cite{Antipov2001}, but is more illustrative of the asymmetry of the load.
%We assume that $u$ decays at infinity; we later evaluate the rate of this decay.
% We seek the solution $u$ with the finite energy
% \begin{equation}
%  \mathcal{E}(u)=\int\limits_{\mathbb{R}^2}|{\nabla u}|^2\mathrm{d}x\mathrm{d}y.
% \end{equation}
% In particular, this gives the conditions
% \begin{equation}\label{plane:zerolimitingtractions}
%  \lim\limits_{y\to+\infty}\mu_1\frac{\partial u}{\partial y}=0,\qquad \lim\limits_{y\to-\infty}\mu_2\frac{\partial u}{\partial y}=0.
% \end{equation}

Near the crack tip, the physical displacement behaves as
\begin{align}
u_j=&\frac{(-1)^{j+1}\sigma_0}{\pi\mu_j}\Biggl\{\frac{\mu_1\kappa\pi}{1+\frac{\mu_1}{\mu_2}}
+\left(1-\ln\left(\frac{r}{b_0}\right)\right)r\cos\theta +(-1)^{j+1}(\pi+(-1)^j\theta)r\sin\theta\Biggr\}+O(r^2\ln^2 r),\quad r\to0,
\end{align}
as demonstrated by \cite{Mish2001a}. It follows that the displacement jump is approximated by
\begin{equation}\label{dispjumpargument}
 \ljump u\rjump(x)=\kappa \sigma_0+O(x\ln |x|),\quad x\to0^\pm,
\end{equation}
as the crack tip is approached along the $x$-axis.

In the neighbourhood of the crack tip, the out of plane component of stress behaves as
\begin{equation}
 \sigma_j\sim\frac{(-1)^j}{\pi}\left\{\sigma_0\ln r\sin\theta+c_0\sin\theta+(-1)^j\sigma_0(\pi+(-1)^j\theta)\cos\theta\right\}
\end{equation}
as $r\to0$, in the usual polar coordinate system and so along the interface,
\begin{equation}\label{plane:whatisa0}
 \sigma\sim \sigma_0,\quad x\to0^+.
\end{equation}
These estimates demonstrate that Fourier transforms of the displacement jump and out-of-plane stress components can be taken; we denote the Fourier transformation $\bar{f}$ of a function $f$ by
\begin{equation}
 \bar{f}(\xi)=\int\limits_{-\infty}^\infty{f(x)e^{i\xi x}\mathrm{d}x}.
\end{equation}
Thus as $\xi\to\infty$, the Fourier transform of the displacement jump behaves as
\begin{equation}\label{plane:uinfinity}
 \ljump \bar{u}\rjump(\xi)=-\kappa \sigma_0 i \xi^{-1} + O(\xi^{-(1+\delta)}),\quad \xi\to\infty,\quad \delta>0.
\end{equation}
Moreover, along the axis, the out of plane stress component decays as
\begin{equation}\label{plane:sigmainfinity}
 \bar{\sigma}=\sigma_0i\xi^{-1}+O(\xi^{-(1+\delta)}),\quad\xi\to\infty,\quad\delta>0.
\end{equation}

\subsection{Weight function formulation}\label{plane:section:weightformulation}

\begin{figure}
 \begin{center}\includegraphics[width=0.6\linewidth]{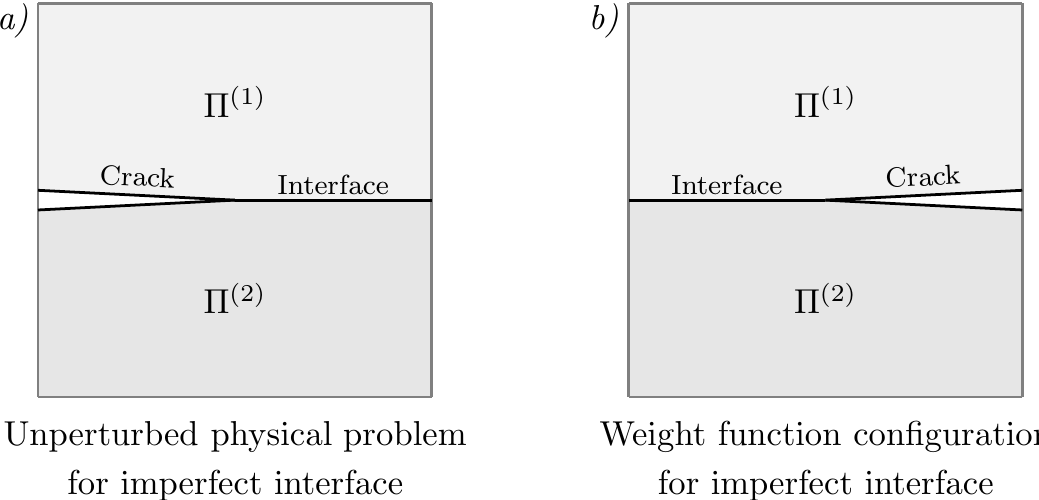}
 \end{center}
 \caption{Geometries for the unperturbed physical (Figure $2a$) and weight function (Figure $2b$) setups.}
\end{figure}

The sought weight function $U$ also satisfies the Laplace equation, but with the crack occupying $\{(x,y):x>0,y=0\}$. We define the functions $\Sigma_j$ in their respective half-planes by
\begin{equation}
 \Sigma_j(x,y):=\mu_j\frac{\partial U_j}{\partial y}, \quad j=1,2.
\end{equation}
Boundary conditions analogous to the physical set-up apply. That is,
\begin{equation}
 \Sigma_1(x,0^+)=\Sigma_2(x,0^-),\quad x\in\mathbb{R},
\end{equation}
\begin{equation}
 \ljump U\rjump(x) -\kappa\Sigma_1(x,0^+)=0,\quad x<0.
\end{equation}
\begin{equation}
 \Sigma_1(x,0^+)=\Sigma_2(x,0^-)=0,\quad x>0.
\end{equation}
We expect that along the interface, the displacement jump behaves as
\begin{align}
\ljump U\rjump(x)&=O(1),\quad x\to0^-;  \\\ljump U\rjump(x)&=O(|x|^{-1/2}),\quad x\to-\infty,
\end{align}
while along the crack, 
\begin{equation}
\ljump U\rjump(x) = c_1+c_2x\log x+c_3x+o(x),\quad x\to0^+, 
\end{equation}
and
\begin{equation}
 \ljump U\rjump(x)=c_4+c_5\sqrt{x}+o(\sqrt{x}),\quad x\to+\infty,
\end{equation}
where $c_i$ are constants. We further expect that
\begin{equation}
 \Sigma_j=O(1),\quad x\to0^-;\qquad \Sigma_j=O(x^{-1/2}),\quad x\to-\infty.
\end{equation}

% \begin{equation}
% U_\pm=\frac{\mp \sigma_0^{(U)}}{\pi\mu_jr}\left\{2\sec\theta\pm\csc\theta\left[4(\pi\mp\theta)\csc^2(2\theta)\mp(\tan\theta-\cot\theta)\right]\right\}+O(1),
% \end{equation}
% $r\to0$, where $\sigma_0^{(U)}$ is an unknown constant to be found from the analysis. Here we have notated $U_1$ and $U_2$ by $U_+$ and $U_-$ respectively due to space constraints. Similarly,
% \begin{equation}
%  \Sigma_j=\frac{(-1)^{j+1}\sigma_0^{(U)}}{\pi r^2}\left\{g(\theta)+\ln r(\csc^3\theta-\sin\theta)\right\}+O\left(\frac{\ln r}{r}\right),\quad r\to0,
% \end{equation}
% where $g(\theta)$ is a known function of $\theta$.

\subsection{Derivation of Wiener-Hopf type equation for the weight function}
The asymptotic behaviour of $U_j$ allows us to apply Fourier transforms. Moreover, the behaviour near $r=0$ demonstrates that the Fourier transform exists as a Cauchy {principal} value integral.
Applying the Fourier transform with respect to $x$
\begin{equation}\label{plane:formofubar}
\bar{U}_j(\xi,y)=\int\limits_{-\infty}^\infty{U_j(x,y)e^{i\xi x}\mathrm{d}x}
\end{equation}
and taking into account the behaviour of $U$ at infinity, we obtain that the transformed solutions of (\ref{plane:laplace}) are of the form
\begin{equation}
 \bar{U}_j(\xi,y)=A_j(\xi)e^{-|\xi y|},
\end{equation}
with the corresponding expressions for tractions at $y=0^\pm$ given by
\begin{equation}
\bar\Sigma_j(\xi,0^\pm)=(-1)^j\mu_j|\xi|A_j(\xi).
\end{equation}
We define the functions $\Phi^\pm(\xi)$ by
\begin{equation}
 \Phi^-(\xi)=\bar{\Sigma}|_{y=0^+},\quad
 \Phi^+(\xi)=\ljump\bar{U}\rjump-\kappa\bar{\Sigma}|_{y=0^+}.
\end{equation}
These functions $\Phi^\pm(\xi)$ are analytic in the complex half planes denoted by their superscripts. We expect that as $\xi\to\infty$ in their respective domains, asymptotic estimates for $\Phi^\pm(\xi)$ are
\begin{equation}\label{plane:phipmatinfinity}
 \Phi^\pm(\xi)=O\left(\frac{1}{\xi}\right),\quad\xi\to\infty,
\end{equation}
and near zero,
\begin{equation}\label{plane:phipmatminusinfinity}
 \Phi^+(\xi)=O(\xi_+^{-3/2}),\qquad \Phi^-(\xi)=O(\xi_-^{-1/2}),\quad\xi\to0;
\end{equation}
we verify this later (see equations (\ref{plane:accuratephibehaviour1})-(\ref{plane:phiplusinfinity})). The condition of continuity of tractions across the crack and interface (\ref{plane:contoftractions}) gives that
\begin{equation}\label{plane:whderivation1}
 \mu_1A_1(\xi)=-\mu_2A_2(\xi),
\end{equation}
% The condition of zero limiting tractions (\ref{plane:zerolimitingtractions}) yields
% \begin{equation}\label{plane:whderivation2}
%  A_j(\xi)\sgn(\xi)+(-1)^{j+1}B_j(\xi)=0,\quad j=1,2.
% \end{equation}
and the Fourier transform of the jump function $\ljump U\rjump$ can be seen from (\ref{plane:formofubar}) to be
\begin{equation}\label{plane:whderivation3}
 \ljump\bar{U}\rjump(\xi)=A_1(\xi)-A_2(\xi).
\end{equation}
Combining these conditions (\ref{plane:whderivation1})-(\ref{plane:whderivation3}), we conclude that the functions $\Phi^\pm(\xi)$ satisfy the functional equation of the Wiener-Hopf type
\begin{equation}\label{plane:whunfactorised}
 \Phi^+(\xi)=-\kappa\Xi(\xi)\Phi^-(\xi),
\end{equation}
where
\begin{equation}\label{plane:Xidefined}
\Xi(\xi)=1+\frac{\mu_0}{|\xi|},
\end{equation}
with the constant $\mu_0$ given by
\begin{equation}
 \mu_0=\frac{\mu_1+\mu_2}{\mu_1\mu_2\kappa}.
\end{equation}
This Wiener-Hopf kernel $\Xi(\xi)$ is the same as that found by \cite{Antipov2001}. The behaviour of the functions $\Phi^\pm(\xi)$ is however different.

\section{Solution of the weight function problem}
\subsection{Factorisation of the weight function Wiener-Hopf kernel}
In this section we factorise the function $\Xi(\xi)$ as defined in (\ref{plane:Xidefined}). As we just remarked, despite this function having been previously factorised in \cite{Antipov2001}, we provide here an alternative factorisation which is more convenient for computations. We define an auxiliary function $\Xi_*$ by
\begin{equation}
 \Xi_*(\xi)=\frac{\xi_+^{1/2}\xi_-^{1/2}}{\xi}\tanh\left(\frac{\xi}{\mu_0}\right)\left(1+\frac{\mu_0}{|\xi|}\right),
\end{equation}
with the functions $\xi_\pm^{1/2}$ given by
\begin{equation}
 \xi_\pm^{1/2}=\sqrt{\mp i\xi}.
\end{equation}
Here ${\sqrt{\cdot}}$ is the standard square root function with its branch cut positioned along the negative real axis. Thus $\xi_\pm^{1/2}$ are analytic functions in half-planes corresponding to their respective subscripts.
Now, $\Xi_*(\xi)$ is an even function and behaves at zero and infinity as follows:
\begin{equation}
 \Xi_*(\xi)=1+\frac{|\xi|}{\mu_0}-\frac{5}{6}\left(\frac{|\xi|}{\mu_0}\right)^2+O\left(\left(\frac{|\xi|}{\mu_0}\right)^3\right),\quad\xi\to0,
\end{equation}
\begin{equation}
 \Xi_*(\xi)=1+\frac{\mu_0}{|\xi|}+O(e^{-2|\xi|/\mu_0}),\quad|\xi|\to\infty.
\end{equation}
The kernel function $\Xi(\xi)$ can be factorised as
\begin{equation}\label{plane:factorisedbigxi}
 \Xi(\xi)=\frac{\xi}{\xi_+^{1/2}\xi_-^{1/2}}\Xi_*(\xi)\Xi_0(\xi),
\qquad
 \Xi_0(\xi)=\coth\left(\frac{\xi}{\mu_0}\right).
\end{equation}
This function can itself be factorised as 
\begin{equation}
\Xi_0(\xi)=\frac{\pi\mu_0}{\xi}\Xi_0^+(\xi)\Xi_0^-(\xi),
\qquad
 \Xi_0^\pm(\xi)=\frac{\Gamma\left(1\mp\frac{i\xi}{\pi\mu_0}\right)}{\Gamma\left(\frac{1}{2}\mp\frac{i\xi}{\pi\mu_0}\right)}.
\end{equation}
The functions $\Xi_0^\pm(\xi)$ satisfy $\Xi_0^+(\xi)=\Xi_0^-(-\xi)$, with $\Xi_0^+(\xi)$ being regular and non-zero in the half plane $\mathrm{Im}(\xi)>-\pi\mu_0/2$. Moreover, Stirling's formula gives that the behaviour as $\xi\to\infty$ in an upper half plane is
\begin{equation}\label{plane:xizeroinfinity}
 \Xi_0^+(\xi)=\beta^{1/2}+\frac{1}{8}\beta^{-1/2}+\frac{1}{128}\beta^{-3/2}+O(\beta^{-5/2}),\quad\xi\to\infty,
\end{equation}
where $\beta=i\xi/(\pi\mu_0)$. Analogous asymptotics for $\Xi_0^-(\xi)$ are easily obtained by noting that $\Xi_0^+(\xi)=\Xi_0^-(-\xi)$.
Near $\xi=0$, the asymptotics for $\Xi_0^+(\xi)$ are given by
\begin{equation}
 \Xi_0^+(\xi)=\frac{1}{\sqrt{\pi}}-\frac{2\ln(2)i\xi}{\pi^{3/2}\mu_0}-\frac{(\pi^2+12\ln^2(2))\xi^2}{6\mu_0^2\pi^{5/2}}+O\left(\xi^3\right),\;\xi\to0.
\end{equation}

The function $\Xi_*(\xi)$ can be written in the form
\begin{equation}
 \Xi_*(\xi)=\Xi_*^+(\xi)\Xi_*^-(\xi),\quad \xi\in\mathbb{C}^\pm
\end{equation}
where
\begin{equation}\label{paper3:xistarpm}
\Xi_*^\pm(\xi)=\exp\left\{{\frac{\pm1}{2\pi i}\int\limits^{\infty}_{-\infty}{\frac{\ln\Xi_*(t)}{t-\xi}\mathrm{d}t}}\right\}.
\end{equation}
% We define an auxiliary function
% \begin{equation}
%  \Theta_*^+(\xi)=\int\limits_{-\infty}^\infty\frac{\ln\Xi_*(t)}{t-\xi}\mathrm{d}t.
% \end{equation}
% We note that since $\Xi_*(t)$ is an even function,
% \begin{equation}
%  \Theta_*^+(0)=0.
% \end{equation}
% Moreover, 
% \begin{equation}
% \Theta_*^+(\xi)=\int\limits_{-\infty}^{\infty}{\left[\frac{\ln\Xi_*(t)}{t-\xi}-\frac{\ln\Xi_*(t)}{t}\right]\mathrm{d}t}=\xi\int\limits_{-\infty}^{\infty}{\frac{\ln\Xi_*(t)}{t(t-\xi)}\mathrm{d}t}\to0\quad\xi\to0;
% \end{equation}
% the integral is bounded. Thus we define
% \begin{equation}
%  \alpha=\int\limits_0^\infty{\frac{\ln\Xi_*(t)}{t^2}}\mathrm{d}t.
% \end{equation}
% It follows that near zero,
% \begin{equation}
%  \Theta_*^+(\xi)=2\alpha\xi+O(\xi^2),\quad\xi\to0,
% \end{equation}
% and so
In particular, we stress that the functions $\Xi_*^\pm(\xi)$ are easy to compute. Near zero, we find that
\begin{equation}\label{plane:xistarplusnear0}
 \Xi_*^+(\xi)=1+\frac{\alpha\xi}{\pi i}+O(\xi^2),\quad\xi\to0,
\end{equation}
where
\begin{equation}
 \alpha=\int\limits_0^\infty{\frac{\ln\Xi_*(t)}{t^2}}\mathrm{d}t,
\end{equation}
which follows from a similar derivation to that of \cite{Vellender2011a}. Moreover, behaviour near infinity in a suitable domain is described by
\begin{equation}\label{plane:xistarinfinity}
\Xi_*^+(\xi)=1+\frac{\mu_0}{\pi i}\frac{\ln(-i\xi)}{\xi}+O\left(\frac{1}{\xi}\right),\quad\text{Im}(\xi)\to+\infty.
\end{equation}

These expressions again emphasise the well behaved nature of the functions $\Xi_*^\pm(\xi)$. The `bad' behaviour of the kernel near $\xi=0$ is all contained in the function $\Xi_0(\xi)$ which has subsequently been factorised into the product of readily computable analytic functions.

\subsection{Solution to the Wiener-Hopf weight function problem}
In this section we solve the Wiener-Hopf problem given in equation (\ref{plane:whunfactorised}).
Substituting our factorised expressions for $\Xi_0^\pm(\xi)$ and $\Xi_*^\pm(\xi)$ into (\ref{plane:whunfactorised}), we arrive at the Wiener-Hopf type equation
\begin{equation}\label{plane:whfactorised}
 \frac{\xi\Phi^+(\xi)\xi_+^{1/2}}{\Xi_0^+(\xi)\Xi_*^+(\xi)}=-\kappa\pi\mu_0\Phi^-(\xi)\Xi_*^-(\xi)\Xi_0^-(\xi)\frac{\xi}{\xi_-^{1/2}}.
\end{equation}
Both sides of (\ref{plane:whfactorised}) represent analytic functions in their respective half-planes and do not have any poles along the real axis. The asymptotic estimates as $\xi\to\infty$ given in (\ref{plane:phipmatinfinity}), (\ref{plane:xizeroinfinity}) and (\ref{plane:xistarinfinity}) demonstrate that both sides of equation (\ref{plane:whfactorised}) behave as $O(1)$ as $\xi\to\infty$ in their respective domains. We therefore deduce that both sides must be equal to a constant, which we denote $\mathcal{A}$.

We deduce that the functions $\Phi^\pm(\xi)$ are given by
\begin{equation}\label{plane:formofphis}
 \Phi^-(\xi)=\frac{-\mathcal{A}\xi_-^{1/2}}{\kappa\pi\mu_0\Xi_*^-(\xi)\Xi_0^-(\xi)\xi},\qquad\Phi^+(\xi)=\frac{\mathcal{A}}{\xi\xi_+^{1/2}}\Xi_0^+(\xi)\Xi_*^+(\xi).
\end{equation}
% and the Fourier transform of $U$ is
% \begin{equation}
%  \bar{U}(\xi,y)=\frac{\mathcal{A}e^{-|\xi y|}}{\kappa\mu_1\Xi_*^+(\xi)\Xi_0^+(\xi)|\xi|\xi_+^{1/2}}.
% \end{equation}%NB THIS NEEDS POSSIBLE MULTIPLICATION BY e^{-|\xi y|}\frac{1}{\mu_1 |\xi|} but somehow needs j dependence. check before reincluding!!
% These expressions verify our earlier claim that $\Phi^\pm(\xi)=O(1/\xi)$ as $\xi\to\infty$. 
These expressions validate our earlier expectations (see equations (\ref{plane:phipmatinfinity}) and (\ref{plane:phipmatminusinfinity}) on page \pageref{plane:phipmatminusinfinity}) regarding the asymptotic estimates for $\Phi^\pm$. In particular, accurate estimates near zero are given by
\begin{equation}\label{plane:accuratephibehaviour1}
 \Phi^-(\xi)=-\frac{\mathcal{A}}{\kappa\mu_0\sqrt{\pi}}\frac{\xi_-^{1/2}}{\xi}\left\{1+\left(\frac{\alpha\mu_0+2\ln2}{\pi i\mu_0}\right)\xi+O(\xi^2)\right\},
\end{equation}
\begin{equation}
 \Phi^+(\xi)=\frac{\mathcal{A}}{\sqrt{\pi}\xi\xi_+^{1/2}}\left\{1+\left(\frac{\alpha\mu_0+2\ln2}{\pi i\mu_0}\right)\xi+O(\xi^2)\right\},\quad\xi\to0,
\end{equation}
while as $\xi\to\infty$ in the appropriate domains,
\begin{equation}\label{plane:phiminusinfinity}
 \Phi^-(\xi)=\frac{-\mathcal{A}}{\xi\kappa\sqrt{\mu_0\pi}}\left(1-\frac{\mu_0}{\pi i}\frac{\ln(-i\xi)}{\xi}+O\left(\frac{1}{\xi}\right)\right),\quad\xi\to\infty,
\end{equation}
\begin{equation}\label{plane:phiplusinfinity}
 \Phi^+(\xi)=\frac{\mathcal{A}}{\xi}\left(\frac{1}{\sqrt{\pi\mu_0}}+\frac{\sqrt{\pi\mu_0}}{i\pi^2}\frac{\ln(-i\xi)}{\xi}+O\left(\frac{1}{\xi}\right)\right),\quad\xi\to\infty.
\end{equation}
It also follows from (\ref{plane:formofphis}) that the Fourier transform of $U$ is given by
\begin{equation}
 \bar{U}_j(\xi,y)=\frac{(-1)^{j+1}\mathcal{A}e^{-|\xi y|}}{\mu_j\kappa\pi\mu_0\Xi_*^-(\xi)\Xi_0^-(\xi)\xi_+^{1/2}\xi},\quad j=1,2.
\end{equation}
% Asymptotic estimates are then easily derived from combining equations (\ref{plane:accuratephibehaviour1})-(\ref{plane:phiplusinfinity}) and noting the relationship
% \begin{equation}
%  \bar{U}_j(\xi,y)=\frac{(-1)^j\Phi^-(\xi)}{\mu_j|\xi|}e^{-|\xi y|}.
% \end{equation}
Expressions for the transforms of the displacement jump and the mean displacement across the interface are therefore respectively given by
\begin{equation}
 \ljump\bar{U}\rjump(\xi)=\frac{\mathcal{A}}{\pi\Xi_*^-(\xi)\Xi_0^-(\xi)\xi_+^{1/2}\xi},
\qquad
 \langle \bar{U} \rangle(\xi):=\frac{1}{2} \left({\bar{U}_1(\xi,0^+)+\bar{U}_2(\xi,0^-)}\right)=\frac{\mathcal{-A\mu_*}}{2\pi\Xi_*^-(\xi)\Xi_0^-(\xi)\xi_+^{1/2}\xi},
\end{equation}
where $\mu_*$ is the dimensionless mechanical contrast parameter 
\begin{equation}
 \mu_*=\frac{(\mu_1-\mu_2)}{(\mu_1+\mu_2)}. 
\end{equation}
These expressions will be useful in Section \ref{plane:section:betti} where we consider the Betti identity in an imperfect interface setting. In particular we note that $\ljump\bar{U}\rjump$ has asymptotic expansions near zero and infinity as follows
\begin{equation}
 \ljump\bar{U}\rjump(\xi)=\frac{\mathcal{A}\sqrt{\pi\mu_0}}{\pi\xi\xi_+^{1/2}\xi_-^{1/2}}\left(1+\frac{\mu_0\ln(i\xi)}{\pi\xi}+O\left(\frac{1}{\xi}\right)\right),\quad \xi\to\infty,
\end{equation}
\begin{equation}
 \ljump\bar{U}\rjump(\xi)=\frac{\mathcal{A}}{\pi^{3/2}\xi\xi_+^{1/2}}\left(1+\frac{(2\ln(2)-\alpha)i\xi}{\pi\mu_0}+O(\xi^2)\right),\quad\xi\to0.
\end{equation}
The function $\langle\bar{U}\rangle$ behaves similarly, as
\begin{equation}
 \langle\bar{U}\rangle(\xi)=\frac{-\mathcal{A}\mu_*\sqrt{\pi\mu_0}}{2\pi\xi\xi_+^{1/2}\xi_-^{1/2}}\left(1+\frac{\mu_0\ln(i\xi)}{\pi\xi}+O\left(\frac{1}{\xi}\right)\right),\quad \xi\to\infty,
\end{equation}
\begin{equation}
 \langle\bar{U}\rangle(\xi)=\frac{-\mathcal{A}\mu_*}{2\pi^{3/2}\xi\xi_+^{1/2}}\left(1+\frac{(2\ln(2)-\alpha)i\xi}{\pi\mu_0}+O(\xi^2)\right),\quad\xi\to0.
\end{equation}

Another key difference between the imperfect and perfect interface (as considered in \cite{PiccWeight}) cases is also readily seen here. Due to the condition of continuity of displacement across perfect interfaces, the function $\jump{\bar U}(\xi)$ is a plus function in the perfect case, since $\jump{U}(x)$ is zero for $x$ lying along the negative real axis. However, across an imperfect interface, the displacement is no longer continuous and so $\jump{\bar U}$ is neither a plus function nor a minus function.

%Combining the expression for $\Phi^+(\xi)$ given in (\ref{plane:formofphis}) with the asymptotic approximations for $\Xi_*^+(\xi)$ and $\Xi_0^+(\xi)$ given in (\ref{plane:xistarinfinity}) and (\ref{plane:xizeroinfinity}) yields that
% \begin{equation}
%  \Phi^+(\xi)=\frac{\mathcal{A}i\sqrt{\mu_0}}{\kappa\xi}\left(1-\frac{\mu_0}{\pi i}\frac{\ln(-i\xi)}{\xi^2}+O(\xi^{-2})\right).
% \end{equation}
 
\section{Betti identity in the imperfect interface setting}\label{plane:section:betti}
In this section we refer to the physical fields for displacement and out-of-plane stress component as $u$ and $\sigma$ respectively, and to the weight function fields for displacement and stress as $U$ and $\Sigma$ respectively. We will use the reciprocal theorem (Betti formula) as in \cite{Willis} to relate the physical solution to the weight function.

Applying the Betti formula to the physical fields and to the upper and lower half plane we obtain
\begin{equation}\label{plane:bettitop}
 \int\limits_{-\infty}^\infty{\left\{U(x'-x,0^+)\sigma(x,0^+)-\Sigma(x'-x,0^+)u(x,0^+)\right\}\mathrm{d}x}=0,
\end{equation}
and
\begin{equation}\label{plane:bettibottom}
 \int\limits_{-\infty}^\infty{\left\{U(x'-x,0^-)\sigma(x,0^-)-\Sigma(x'-x,0^-)u(x,0^-)\right\}\mathrm{d}x}=0.
\end{equation}
These identities were proved under the assumption that the integrand decays faster at infinity than $1/R$ along any ray. It is clear from the asymptotic estimates for the physical solution and the weight function given in subsections \ref{plane:section:physicalformulation} and \ref{plane:section:weightformulation} that this condition is satisfied. Subtracting (\ref{plane:bettibottom}) from (\ref{plane:bettitop}) we obtain
\begin{align}
  \int\limits_{-\infty}^\infty\{&U(x'-x,0^+)\sigma(x,0^+)-U(x'-x,0^-)\sigma(x,0^-)\nonumber\\
&-[\Sigma(x'-x,0^+)u(x,0^+)-\Sigma(x'-x,0^-)u(x,0^-)]\}\mathrm{d}x=0.\label{plane:bettishort}
\end{align}
We split the terms for physical stress into two parts, writing
\begin{equation}\label{plane:splitting}
 \sigma(x,0^\pm)=p_\pm^{(-)}(x)+\sigma^{(+)}(x),
\end{equation}
where $p_\pm^{(-)}$ and $\sigma^{(+)}$ are defined as follows
\begin{equation}
 p_\pm^{(-)}(x)=H(-x)\sigma(x,0^\pm),\qquad
\sigma^{(+)}(x)=H(x)\sigma(x,0);
\end{equation}
here $H(x)$ denotes the Heaviside step function. The functions $p^\pm$ represent the prescribed loading on the crack faces. After this splitting, equation (\ref{plane:bettishort}) becomes
\begin{align}
 \int\limits_{-\infty}^\infty\{&\ljump U\rjump(x'-x)\sigma^{(+)}(x)-\Sigma(x'-x,0)\ljump u\rjump(x)\}\mathrm{d}x=-\int\limits_{-\infty}^\infty\{U(x'-x,0^+)p^+(x)-U(x'-x,0^-)p^-(x)\}\mathrm{d}x.\label{plane:bettilong}
\end{align}
%%%%%!
We introduce notation for symmetric and skew-symmetric parts of the loading:
\begin{equation}\label{plane:unpertloading}
 \langle p\rangle^{(-)}(x)=\frac{1}{2}(p_+^{(-)}(x)+p_-^{(-)}(x)),\quad\ljump p\rjump^{(-)}(x)=p_+^{(-)}(x)-p_-^{(-)}(x).
\end{equation}

This allows us to rewrite the right hand side of (\ref{plane:bettilong}), giving
\begin{align}
 \int\limits_{-\infty}^\infty\{&\ljump U\rjump(x'-x)\sigma^{(+)}(x)-\Sigma(x'-x,0)\ljump u\rjump(x)\}\mathrm{d}x\nonumber
\\&=-\int\limits_{-\infty}^\infty\{\ljump U\rjump(x'-x)\langle p\rangle(x)+\langle U\rangle(x'-x)\ljump p\rjump(x)\}\mathrm{d}x.\label{plane:bettilong2}
\end{align}
We now split $\ljump U \rjump$ into the sum of $\ljump U \rjump^{(\pm)}$ in the spirit of (\ref{plane:splitting}), and similarly split $\ljump{u}\rjump$ into the sum of $\ljump{u}\rjump^{(\pm)}$. We will use the usual notation of $f*g$ to denote the convolution of $f$ and $g$. Rewriting (\ref{plane:bettilong2}) using these expressions gives
\begin{align}
 \ljump U\rjump^{(+)}*\sigma^{(+)}+\ljump U\rjump^{(-)}*\sigma^{(+)}&-\Sigma*\ljump u\rjump^{(+)}-\Sigma*\ljump u\rjump^{(-)}=-\ljump U\rjump*\langle p\rangle^{(-)}-\langle U\rangle*\ljump p\rjump^{(-)}.
\end{align}
Taking Fourier transforms in $x$ yields
\begin{align}
 \overline{\ljump U\rjump^{(+)}}(\xi)&\overline{\sigma^{(+)}}(\xi)+\overline{\ljump U\rjump^{(-)}}(\xi)\overline{\sigma^{(+)}}(\xi)-\overline{\Sigma}(\xi)\overline{\ljump u\rjump^{(+)}}(\xi)-\overline{\Sigma}(\xi)\overline{\ljump u\rjump^{(-)}}(\xi)=-\bar{\ljump U\rjump}(\xi)\bar{\langle p\rangle}(\xi)-\bar{\langle U\rangle}(\xi)\bar{\ljump p\rjump}(\xi).\label{plane:bettilong3}
\end{align}
We now make use of the transmission conditions which state that
\begin{equation}
 \overline{\ljump{U}\rjump^{(-)}}(\xi)=\kappa\overline{\Sigma}(\xi),\qquad \overline{\ljump{u}\rjump^{(+)}}(\xi)=\kappa\overline{\sigma^{(+)}}(\xi).
\end{equation}
This causes the second and third terms in the left hand side of (\ref{plane:bettilong3}) to cancel, leaving
\begin{align}
 \overline{\ljump U\rjump^{(+)}}(\xi)&\overline{\sigma^{(+)}}(\xi)-\overline{\Sigma}(\xi)\overline{\ljump u\rjump^{(-)}}(\xi)=-\bar{\ljump U\rjump}(\xi)\bar{\langle p\rangle}(\xi)-\bar{\langle U\rangle}(\xi)\bar{\ljump p\rjump}(\xi),\quad \xi\in\mathbb{R}.\label{plane:betti}
\end{align}
We note that
\begin{equation}
 \overline{\ljump {U}\rjump^{(+)}}(\xi)\equiv\Phi^+(\xi),\qquad \bar{\Sigma}(\xi)\equiv\Phi^-(\xi),
\end{equation}
and can therefore combine the asymptotic estimates in (\ref{plane:uinfinity}), (\ref{plane:sigmainfinity}), (\ref{plane:phiminusinfinity}) and (\ref{plane:phiplusinfinity}) to yield that
\begin{align}
\overline{\ljump U\rjump^{(+)}}(\xi)\overline{\sigma^{(+)}}(\xi)=\frac{\sigma_0i}{\sqrt{\pi\mu_0}\xi^2}+O(\xi^{-(2+\delta)}),\qquad\xi\to\infty\text{ in }\mathbb{C}^+,\\ 
\bar\Sigma(\xi)\overline{\ljump u\rjump^{(-)}}(\xi)=\frac{\sigma_0i}{\sqrt{\pi\mu_0}\xi^2}+O(\xi^{-(2+\delta)}),\qquad\xi\to\infty\text{ in }\mathbb{C}^-,
\end{align}
where $\delta>0$. We now multiply both sides of (\ref{plane:betti}) by $\xi$, giving
\begin{align}
 \xi&\left(\overline{\ljump U\rjump^{(+)}}(\xi)\overline{\sigma^{(+)}}(\xi)-\overline{\Sigma}(\xi)\overline{\ljump u\rjump^{(-)}}(\xi)\right)=-\xi\left(\bar{\ljump U\rjump}(\xi)\bar{\langle p\rangle}(\xi)+\bar{\langle U\rangle}(\xi)\bar{\ljump p\rjump}(\xi)\right).\label{plane:bettibyxi}
\end{align}
Then, and similarly to the expression obtained for the perfect interface Betti formula approach of \cite{Willis}, the left hand side now has asymptotics at infinity (in appropriate domains) of the form
\begin{equation}
 \xi\left(\overline{\ljump U\rjump^{(+)}}(\xi)\overline{\sigma^{(+)}}(\xi)-\overline{\Sigma}(\xi)\overline{\ljump u\rjump^{(-)}}(\xi)\right)=\frac{\sigma_0i}{\sqrt{\pi\mu_0}}\left[\frac{1}{\xi+i0}-\frac{1}{\xi-i0}\right]
\end{equation}
as $\xi\to\infty$, where the term in square brackets is the regularization of the Dirac delta function, namely $-2\pi i\delta(\xi)$. Integrating both sides of (\ref{plane:bettibyxi}), we can arrive {at} an expression for the constant $\sigma_0$ in terms of known, readily computable functions:
\begin{equation}\label{plane:expressionfora0}
 {\sigma_0}=\frac{1}{2}\sqrt{\frac{\mu_0}{\pi}}\int\limits_{-\infty}^\infty\xi\left(\bar{\ljump U\rjump}(\xi)\bar{\langle p\rangle}(\xi)+\bar{\langle U\rangle}(\xi)\bar{\ljump p\rjump}(\xi)\right)\mathrm{d}\xi.
\end{equation}
We note that since $\ljump\bar{U}\rjump(\xi)$ and $\langle\bar{U}\rangle(\xi)$ behave as $O(\xi^{-2})$ as $\xi\to\infty$, and the functions $\ljump\bar{p}\rjump(\xi)$ and $\langle\bar{p}\rangle(\xi)$ behave as bounded oscillations as $\xi\to\infty$ for point loadings, the integrand is well behaved at infinity. Moreover, near $\xi=0$ the integrand is also sufficiently well behaved, acting as $O(\xi_+^{-1/2})$.

Equation (\ref{plane:expressionfora0}) is a particularly important result; it gives an expression for the leading order of the out-of-plane component of stress near the crack tip (see (\ref{plane:whatisa0})) in terms of known functions and acts as an imperfect interface analogue to the stress intensity factor from the perfect interface setting.

\subsection{The functions $\jump{\bar p}$ and $\av{\bar{p}}$ for specific point loadings}
As we stated earlier in this paper, although the methods described are applicable to any permissible loading, we will later perform computations using the specific point loading configuration shown in Figure \ref{plane:figure:physicaldefectsetup} on page \pageref{plane:figure:physicaldefectsetup}.

For this configuration, the loadings are defined as a point load on the upper crack face at $x=-a$ balanced by two equal loads at $x=-a+b$ and $x=-a-b$, that is
\begin{equation}\label{pointloadingsdefn}
 p_+^{(-)}(x)=F\delta(x+a),\quad p_-^{(-)}(x)=\frac{F}{2}\left(\delta(x+a+b)+\delta(x+a-b)\right).
\end{equation}
The corresponding explicit expressions for $\langle p\rangle(x)$ and $\ljump p\rjump(x)$ are
\begin{align}
 \langle p\rangle(x)&=\frac{F}{2}\left\{\delta(x+a)+\frac{1}{2}\left(\delta(x+a+b)+\delta(x+a-b)\right)\right\},\label{plane:pmean}\\
 \ljump p\rjump(x)&=F\left\{\delta(x+a)-\frac{1}{2}\left(\delta(x+a+b)+\delta(x+a-b)\right)\right\}\label{plane:pjump},
\end{align}
which have Fourier transforms given by 
% For our specific point force loading, we note that the expressions for $\langle p\rangle(x)$ and $\ljump p\rjump(x)$ in (\ref{plane:pmean}) and (\ref{plane:pjump}) imply that the transformed functions $\langle \bar{p}\rangle(\xi)$ and $\ljump{\bar{p}}\rjump(\xi)$ are given by
\begin{align}
 \langle\bar{p}\rangle(\xi)&=\frac{F}{4}(e^{ib\xi}+1)^2e^{-i(a+b)\xi},\quad
 \ljump{\bar{p}}\rjump(\xi)=-\frac{F}{2}(e^{ib\xi}-1)^2e^{-i(a+b)\xi}.
\end{align}

\section{The unperturbed solution, $u_0$}
We will later require a method to evaluate the {\em unperturbed physical solution} $u_0$ and its first order partial derivatives with respect to $x$ and $y$. This problem has been solved by \cite{Antipov2001} by approximating the loading by a linear combination of exponentials; this approximation is however not ideal for point loadings.

% In this section we consider point loadings applied to the crack faces. The loading is split into symmetric and anti-symmetric parts; we consider these separately. Since the problem is linear, we may always consider such a loading by decomposing it into the sum of symmetric and anti-symmetric loadings.

Tractions on the upper and lower crack faces can be written as
\begin{equation}
 \bar{\sigma}_1(\xi,0^+)=\bar{p}_1(\xi)+\varphi_1^+(\xi),\quad
 \bar{\sigma}_2(\xi,0^-)=\bar{p}_2(\xi)+\varphi_2^+(\xi).
\end{equation}
It follows immediately from continuity of tractions across the imperfect interface that
\begin{equation}
 \varphi_1^+(\xi)=\varphi_2^+(\xi)=:\varphi^+(\xi).
\end{equation}
We further define minus functions, $\varphi_1^-$ and $\varphi_2^-$ as
\begin{equation}
 \varphi_1^-(\xi)=\ljump\bar{u}\rjump(\xi)-\kappa\bar\sigma_1(\xi,0^+),
\quad
 \varphi_2^-(\xi)=\ljump\bar{u}\rjump(\xi)-\kappa\bar\sigma_2(\xi,0^-).
\end{equation}
We expect that the unknown functions $\varphi^+(\xi)$ and $\varphi_j^-(\xi)$ behave at infinity as
\begin{equation}
 \varphi_j^\pm(\xi)=O\left(\frac{1}{\xi}\right),\quad\xi\to\infty,\quad\pm\mathrm{Im}(\xi)>0.
\end{equation}
% \begin{equation}
%  \varphi_j^-(\xi)=O\left(\frac{1}{\xi}\right),\quad\xi\to\infty.
% \end{equation}
From these expressions follow the relationships
\begin{equation}
 \ljump\bar{\sigma}\rjump(\xi)\equiv\ljump\bar p\rjump(\xi),
\end{equation}
\begin{equation}
 \langle\bar{\sigma}\rangle(\xi)\equiv\langle \bar{p}\rangle(\xi)+\varphi^+(\xi),
\end{equation}
and also
\begin{equation}
\label{plane:3rdcond}
 -\kappa\ljump\bar{\sigma}\rjump(\xi)\equiv\varphi_1^-(\xi)-\varphi_2^-(\xi),
\end{equation}
\begin{equation}
\label{plane:4thcond}
 2\ljump\bar{u}\rjump(\xi)-2\kappa\langle\bar{\sigma}\rangle(\xi)\equiv\varphi_1^-(\xi)+\varphi_2^-(\xi).
\end{equation}
Moreover, since transformed solutions are of the form
\begin{equation}\label{plane:whatareaj}
 \bar{u}_j(\xi,y)=A_j(\xi)e^{-|\xi y|},
\end{equation}
we further have the relationships
\begin{equation}
 \ljump\bar{u}\rjump(\xi)=A_1(\xi)-A_2(\xi),
\end{equation}
\begin{equation}
 \ljump\bar{\sigma}\rjump(\xi)=-|\xi|(\mu_1A_1(\xi)+\mu_2A_2(\xi)),
\end{equation}
and
\begin{equation}
 \langle\bar{\sigma}\rangle(\xi)=\frac{|\xi|}{2}(\mu_2A_2(\xi)-\mu_1A_1(\xi)).
\end{equation}
These seven equations in eight unknowns reduce to the following Wiener-Hopf type equation relating $\varphi^+(\xi)$ and $\varphi_1^-(\xi)$:
\begin{align}
 -\kappa\left\{1+\frac{\mu_0}{|\xi|}\right\}&\varphi^+(\xi)-\kappa\left\{1+\frac{\mu_0}{|\xi|}\right\}\av{\overline{p}}(\xi)=\varphi_1^-(\xi)+\frac{\kappa}{2}\left\{1-\frac{\mu_*\mu_0}{|\xi|}\right\}\jump{\overline{p}}(\xi).\label{plane:antiwh}
\end{align}
Noting that the term in braces on the left hand side of (\ref{plane:antiwh}) is the function we earlier defined as $\Xi(\xi)$ and have already suitably factorised, we can write
\begin{equation}
 -\kappa\Xi(\xi)\varphi^+(\xi)-\kappa\Xi(\xi)\av{\overline{p}}(\xi)=\varphi_1^-(\xi)+\kappa\Lambda(\xi)\jump{\overline{p}}(\xi),
\end{equation}
 where
\begin{equation}
 \Lambda(\xi)=\frac{1}{2}\left\{1-\frac{\mu_*\mu_0}{|\xi|}\right\}.
\end{equation}
Recall that $\Xi(\xi)$ can be factorised in the form
\begin{equation}
 \Xi(\xi)=\pi\mu_0B^+(\xi)B^-(\xi),
\end{equation}
where we have defined the functions $B^\pm(\xi)$ for the sake of notational brevity by
\begin{equation}
 B^+(\xi)=\frac{\Xi_0^+(\xi)\Xi_*^+(\xi)}{\xi_+^{1/2}},\qquad B^-(\xi)=\frac{\Xi_0^-(\xi)\Xi_*^-(\xi)}{\xi_-^{1/2}},
\end{equation}
which are analytic in the half planes indicated by their superscripts. These functions have behaviour near zero and infinity given by 
\begin{equation}\label{plane:bpmzero}
    B^\pm(\xi)=O(\xi^{-1/2}),\qquad \xi\to0,                                                                                                                 
\end{equation}
\begin{equation}\label{plane:bpminf}
    B^\pm(\xi)=O(1),\qquad\xi\to\infty.
\end{equation}
Thus
\begin{equation}
-\kappa\pi\mu_0B^+(\xi)\varphi^+(\xi)=
\frac{\varphi_1^-(\xi)}{B^-(\xi)}+\kappa\frac{\Lambda(\xi)}{B^-(\xi)}\jump{\overline{p}}(\xi)+\kappa\pi\mu_0B^+(\xi)\av{\overline{p}}(\xi).
\end{equation}
We can decompose the final term on the right hand side as usual into
\begin{equation}
\kappa\frac{\Lambda(\xi)}{B^-(\xi)}\jump{\overline{p}}(\xi)+\kappa\pi\mu_0B^+(\xi)\av{\overline{p}}(\xi)=L^+(\xi)-L^-(\xi),
\end{equation}
where $L^\pm(\xi)$ are given by
\begin{equation}\label{plane:ldefinition}
L^\pm(\xi)=\frac{1}{2\pi i}\int\limits_{-\infty}^\infty\left\{\kappa\frac{\Lambda(\beta)}{B^-(\beta)}\jump{\overline{p}}(\beta)+\kappa\pi\mu_0B^+(\beta)\av{\overline{p}}(\beta)\right\}
\frac{\mathrm{d}\beta}{\beta-\xi},
\end{equation}
for $\quad\xi\in\mathbb{C}^\pm$. We expect that $L^\pm(\xi)$ behave as $O(\xi^{-1})$ as $\xi\to\infty$.

The Wiener-Hopf equation becomes
\begin{equation}\label{plane:antisymmwh}
-\kappa\pi\mu_0B^+(\xi)\varphi^+(\xi)-L^+(\xi)=
\frac{\varphi_1^-(\xi)}{B^-(\xi)}-L^-(\xi).
\end{equation}
Both terms on each side of (\ref{plane:antisymmwh}) decay as $O(1/\xi)$, $\xi\to\infty$. Moreover, each side is analytic in the half-plane denoted by the superscripts. Liouville's theorem yields that both sides are equal to zero, and so
\begin{equation}\label{plane:phiplusexpression}
 \varphi^+(\xi)=\frac{-L^+(\xi)}{\kappa\pi\mu_0B^+(\xi)},
\qquad
 \varphi_1^-(\xi)=L^-(\xi)B^-(\xi).
\end{equation}
These expressions verify that our expectations of the behaviour of $\varphi^+(\xi)$ and $\varphi_1^-(\xi)$ as $\xi\to\infty$ were correct.
Moreover, (\ref{plane:3rdcond}) enables us to express $\varphi_2^-(\xi)$ as
\begin{equation}
 \varphi_2^-(\xi)=L^-(\xi)B^-(\xi)+\kappa\ljump\bar p\rjump(\xi).
\end{equation}
Condition (\ref{plane:4thcond}) then yields an expression for the transform of the displacement jump
\begin{align}
\ljump\bar{u}\rjump(\xi)&=\frac{\varphi_1^-(\xi)}{2}+\frac{\varphi_2^-(\xi)}{2}+\kappa\av{\bar \sigma}(\xi)=
\varphi_1^-(\xi)+\kappa\varphi^+(\xi)+\kappa\av{\bar p}(\xi)+\frac{\kappa}{2}\jump{\bar p}(\xi),
\end{align}
from which we can obtain expressions for $A_1(\xi)$ and $A_2(\xi)$ as follows
\begin{equation}\label{plane:A1formula}
 A_1(\xi)=-\frac{1}{\mu_1|\xi|}\left\{\varphi^+(\xi)+\av{\bar p}(\xi)+\frac{1}{2}\jump{\bar p}(\xi)\right\},
\quad
 A_2(\xi)=\frac{1}{\mu_2|\xi|}\left\{\varphi^+(\xi)+\av{\bar p}(\xi)-\frac{1}{2}\jump{\bar p}(\xi)\right\}.
\end{equation}
These expressions now enable us (see (\ref{plane:whatareaj})) to compute the Fourier transform of the unperturbed solution (i.e. the setup with no small defect present) $\bar u_j(\xi,y)$ for any $\xi,y$.

\section{Perturbation analysis}\label{section:pert}

We shall construct an asymptotic solution of the problem using the method of \cite{Movchanmovchan}, that is the asymptotics of the solution will 
be taken in the form
\begin{equation}
\label{exp}
u_{1,2}(\bx,\varepsilon) = u_{1,2}^{(0)}(\bx)  + \varepsilon W^{(1)}(\bxi) + \varepsilon^2 u_{1,2}^{(1)}(\bx) + o(\varepsilon^2), \quad \varepsilon \to 0. 
\end{equation}
In (\ref{exp}), the leading term $u_{1,2}^{(0)}(\bx)$ corresponds to the unperturbed solution, which is described in the previous section. {The small dimensionless parameter $\varepsilon$ is defined as the ratio of the semi-major axis of the elliptical inclusion to the distance of the defect's center from the crack tip, that is $\varepsilon=l/d$.} The term 
$\varepsilon W^{(1)}(\bxi)$ corresponds to the boundary layer concentrated near the defect and needed to satisfy the transmission conditions 
for the elastic inclusion $g_\varepsilon$
\begin{equation}
\label{inclusion} 
u_{\text{in}} = u_{\text{out}}, \quad \mu_{\text{in}} \frac{\partial u_{\text{in}}}{\partial n} = \mu_{\text{out}} \frac{\partial u_{\text{out}}}{\partial n} \quad 
\text{on $\partial g_\varepsilon$}.
\end{equation}

The term $\varepsilon^2 u_{1,2}^{(1)}(\bx)$ is introduced to fulfil the original boundary 
conditions (4) on the crack faces and the interface conditions (2), (3) disturbed by the boundary layer; this term, in turn, will produce 
perturbations of the crack tip fields and correspondingly of the constant $\sigma_0$.

We shall consider an elastic inclusion, situated in the upper (or lower) half-plane. The leading term $u_{1,2}^{(0)}$ clearly does not satisfy the transmission 
conditions (\ref{inclusion}) on the boundary $\partial g_\varepsilon$. Thus, we shall correct the solution by constructing the boundary layer $W^{(1)}(\bxi)$, 
where the new scaled variable $\bxi$ is defined by
\begin{equation}
\label{Y1}
\bxi = \frac{\bx - \bY}{\varepsilon},
\end{equation}
with $\bY = (X,Y)$ being the ``centre'' of the inclusion $g_\varepsilon$ (see Figure \ref{plane:figure:physicaldefectsetup}).

For $W^{(1)}(\bxi) = \{W^{(1)}_{\text{in}}, \ \bxi \in g; \ W^{(1)}_{\text{out}}, \ \bxi \in \mathbb{R}^2 \setminus \overline{g}\}$ we consider the following problem 
\begin{equation}
\label{laplace1} 
\nabla^2 W^{(1)}_{\text{in}}(\bxi) = 0, \quad \bxi \in g, \quad \nabla^2 W^{(1)}_{\text{out}}(\bxi) = 0, \quad \bxi \in \mathbb{R}^2 \setminus \overline{g},
\end{equation}
where
$$
g = \varepsilon^{-1} g_\varepsilon \equiv \{\bxi \in \mathbb{R}^2: \varepsilon\bxi + \bY \in g_\varepsilon\}.
$$
The function $W^{(1)}$ remains continuous across the interface $\partial g$, that is, 
$$
W^{(1)}_{\text{in}} = W^{(1)}_{\text{out}} \quad \text{on} \quad \partial g,
$$
and satisfies on $\partial g$ the following transmission condition 
\begin{equation}
\mu_{\text{in}} \frac{\partial}{\partial \bn} W^{(1)}_{\text{in}}(\bxi) - \mu_{\text{out}} \frac{\partial}{\partial \bn} W^{(1)}_{\text{out}}(\bxi) = 
(\mu_{\text{out}} - \mu_{\text{in}}) \bn \cdot \nabla u^{(0)}(\bY) + O(\varepsilon),
\end{equation}
as $\varepsilon \to 0$, where $\bn = \bn_{\bxi}$ is an outward unit normal on $\partial g$. The formulation is completed by setting the following condition at infinity
\begin{equation}
\label{infinity}
W^{(1)}_{\text{out}} \to 0 \quad \text{as} \quad |\bxi| \to \infty.
\end{equation}
The problem above has been solved by various techniques and the solution can be found, for example, in  \cite{Movchanmovchan}.

Since we assume that the inclusion is at a finite distance from the interface between the half-planes, we shall only need the leading term of the 
asymptotics of the solution at infinity. This term reads as follows
\begin{equation}
\label{boundary_layer_1}
W^{(1)}_{\text{out}}(\bxi) = -\frac{1}{2\pi} \left[\left.\nabla_{\bx}\, u^{(0)}\right|_{\bY}\right] \cdot \left[\bmM \frac{\bxi}{|\bxi|^2}\right] + 
O(|\bxi|^{-2}) \quad \text{as} \quad \bxi \to \infty, 
\end{equation}
where $\bmM$ is a 2 $\times$ 2 matrix which depends on the characteristic size $\ell$ of the domain $g$ and the ratio $\mu_{\text{out}}/\mu_{\text{in}}$; it is called the dipole 
matrix. For example, in the case of an elliptic inclusion with the semi-axes $\ell_a$ and $\ell_b$ making an angle $\alpha$ with the positive direction of the $x$-axis 
and $y$-axis, respectively, the matrix $\bmM$ takes the form
\begin{equation}
\label{dip}
\bmM = -\frac{\pi}{2} \ell_a \ell_b (1 + e)({\nu_*} - 1)\mathcal{B},
\end{equation}
where
\begin{equation}
\mathcal{B}=\left[
\begin{array}{cc}
\displaystyle 
\frac{1 + \cos 2\alpha}{e + {\nu_*}} + \frac{1 - \cos 2\alpha}{1 + e{\nu_*}} &
\displaystyle 
-\frac{(1 - e)({\nu_*} - 1) \sin 2\alpha}{(e + {\nu_*})(1 + e{\nu_*})} \\[3mm]
\displaystyle 
-\frac{(1 - e)({\nu_*} - 1) \sin 2\alpha}{(e + {\nu_*})(1 + e{\nu_*})} & 
\displaystyle 
\frac{1 - \cos 2\alpha}{e + {\nu_*}} + \frac{1 + \cos 2\alpha}{1 + e{\nu_*}} 
\end{array}
\right],
\end{equation}
$e = \ell_b/\ell_a$ and ${\nu_*} = \mu_{\text{out}}/\mu_{\text{in}}$. We note that for a soft inclusion, $\mu_{\text{out}} > \mu_{\text{in}}$, the dipole matrix is negative 
definite, whereas for a stiff inclusion, $\mu_{\text{out}} < \mu_{\text{in}}$, the dipole matrix is positive definite.
In the limit $\mu_{\text{in}} \to \infty$, we obtain the dipole matrix for a rigid movable inclusion. In the case of an elliptic rigid inclusion, we have
\begin{equation}
\label{rigid-ellipse}
\bmM_{\mathrm{rig}} = \frac{\pi}{2} \ell_a \ell_b (1/e + 1) \mathcal{B}_{\mathrm{rig}},
\end{equation}
where
\begin{equation}
\mathcal{B}_{\mathrm{rig}}=\left[
\begin{array}{cc}
h_+(\alpha)+e h_-(\alpha) & (1 - e)\sin 2\alpha \\[3mm]
(1 - e)\sin 2\alpha & h_-(\alpha)+e h_+(\alpha)
\end{array}
\right].
\end{equation}
Here we have defined the functions $h_\pm(\alpha)=1\pm\cos2\alpha$ for brevity of notation. The term $\varepsilon W^{(1)}(\bxi)$ in a neighbourhood of the $x$-axis written in the $\bx$ coordinates takes the form
\begin{equation}
\label{BL_1_crack_faces_1} 
\varepsilon W^{(1)}(\bxi) = \varepsilon^2 w^{(1)}(\bx) + o(\varepsilon^2), \quad \varepsilon \to 0,
\end{equation}
where
\begin{equation}
\label{BL_1_crack_faces_2} 
w^{(1)}(\bx) = -\frac{1}{2\pi} \left[\left.\nabla_{\bx} u^{(0)}\right|_{\bY}\right] \cdot \left[\bmM \frac{\bx - \bY}{|\bx - \bY|^2}\right]. 
\end{equation}

As a result, one can compute the average $\varepsilon^2 \langle \sigma^{(1)} \rangle$ and the jump $\varepsilon^2 \jump{\sigma^{(1)}}$ of the ``effective'' tractions 
on the crack faces induced by the elastic inclusion $g_\varepsilon$. 
Since $\partial u_{1,2}^{(1)}/\partial y = -\partial w^{(1)}/\partial y$ must hold on the crack line (to satisfy the original boundary conditions (4)), this gives for $x < 0$
\begin{equation}
\label{traction_discrepancy_1} 
\langle \sigma^{(1)} \rangle^{(-)}(x) = 
-\frac{1}{2}(\mu_1 + \mu_2) \frac{\partial w^{(1)}}{\partial y} := P^{(-)}(x), 
\end{equation}
\begin{equation}
\jump{\sigma^{(1)}}^{(-)}(x) = 
-(\mu_1 - \mu_2) \frac{\partial w^{(1)}}{\partial y} := Q^{(-)}(x), 
\end{equation}
where
\begin{align}
\label{dw1dx2}
\frac{\partial w^{(1)}}{\partial y} = 
-\frac{1}{2\pi} &\left[\left.\nabla_{\bx} u^{(0)}\right|_{\bY}\right] \cdot \bmM \frac{\be_2}{|\bx - \bY|^2}+ 
\frac{1}{\pi} \left[\left.\nabla_{\bx} u^{(0)}\right|_{\bY}\right] \cdot \bmM \frac{(\bx - \bY)(y - Y)}{|\bx - \bY|^4}.
\end{align}
Additionally, we can compute the transmission conditions for the functions $u_{1,2}^{(1)}$ across the interface. In order for the perturbed solution $u_{1,2}$ in (\ref{exp}) to 
satisfy the original transmission conditions (2) and (3), the following relations must hold for $x > 0$
\begin{align}
\jump{u^{(1)}}^{(+)}(x) &= \kappa \av{\sigma^{(1)}}^{(+)}(x) + \frac{\kappa}{2}(\mu_1 + \mu_2) \frac{\partial w^{(1)}}{\partial y}\nonumber \\&:= 
\kappa \av{\sigma^{(1)}}^{(+)}(x) - \kappa P^{(+)}(x), 
\end{align}
\begin{equation}
\jump{\sigma^{(1)}}^{(+)}(x) = -(\mu_1 - \mu_2) \frac{\partial w^{(1)}}{\partial y} := Q^{(+)}(x).
\end{equation}
%Note that we have defined the functions $P^{(\pm)}$ and $Q^{(\pm)}$ above.
% We will later need an expression for the Fourier transform of $\partial w^{(1)}/\partial y$, so we note the useful expressions for $\xi\in\mathbb{R}$
% \begin{equation}
%  \int\limits_{-\infty}^\infty{\frac{e^{i\xi x}\mathrm{d}x}{|\bx-\bY|^2}}=\int\limits_{-\infty}^\infty{\frac{e^{i\xi x}\mathrm{d}x}{(x-X)^2+Y^2}}=\frac{\pi}{Y} e^{i\xi X}e^{-Y|\xi|},
% \end{equation}
% and
% \begin{equation}
%  \int\limits_{-\infty}^\infty{\frac{e^{i\xi x}\mathrm{d}x}{|\bx-\bY|^4}}=\int\limits_{-\infty}^\infty{\frac{e^{i\xi x}\mathrm{d}x}{\left((x-X)^2+Y^2\right)^2}}=\frac{\pi}{2|Y|^3}e^{i\xi X}e^{-|\xi Y|}(1+|\xi Y|).
% \end{equation}

\section{Model problem for the first order perturbation}

The constant $\sigma_0$ which describes the traction near the crack tip (see (\ref{plane:whatisa0})) is expanded in the form
\begin{equation}
\sigma_0 = \sigma_0^{(0)} + \varepsilon^2 \Delta \sigma_0 + o(\varepsilon^2), \quad \varepsilon \to 0.
\end{equation}
Our objective is to find the first order variation $\Delta \sigma_0$.

Let us consider the model problem for the first order perturbation $u^{(1)}$ and write the corresponding Betti identity in the form
\begin{align}
\int_{-\infty}^{\infty} \biggl\{\jump{U}(x' - x) \av{\sigma^{(1)}}(x) &+ \av{U}(x' - x)\jump{\sigma^{(1)}}(x) - \av{\Sigma}(x' - x)\jump{u^{(1)}}(x)\biggr\} \mathrm{d}x = 0.\label{betti}
\end{align}
This follows immediately from (\ref{plane:bettishort}) by noting that $\jump{\Sigma}\equiv 0$. We split the terms for stress into two parts,
\begin{equation}
\av{\sigma^{(1)}} = \av{\sigma^{(1)}}^{(+)} + P^{(-)}, \qquad 
\jump{\sigma^{(1)}} = Q^{(+)} + Q^{(-)},
\end{equation}
observing that in contrast to the zero order problem where the load is described by (\ref{plane:unpertloading}), the terms with superscript $^{(+)}$ are non-zero since the presence of inclusions induces stresses along the imperfect interface and should be taken into account. Equation (\ref{betti}) becomes
\begin{equation}
\begin{array}{l}
\displaystyle
\int_{-\infty}^{\infty} \left\{ \jump{U}(x' - x) \av{\sigma^{(1)}}^{(+)}(x) - \av{\Sigma}(x' - x)\jump{u^{(1)}}(x) \right\} \mathrm{d}x  \\[3mm]
\displaystyle
\quad=-\int_{-\infty}^{\infty} \Biggl\{ \jump{U}(x' - x) P^{(-)}(x) + \av{U}(x' - x)Q^{(-)}(x)+ \av{U}(x' - x)Q^{(+)}(x) \Biggr\} \mathrm{d}x.
\end{array}
\end{equation}
We now split $\jump{U}$ into the sum of $\jump{U}^{\pm}$ and similarly split $\jump{u}$ into the sum of $\jump{u}^{\pm}$. This gives
\begin{align}
\jump{U}^{(+)} &* \av{\sigma^{(1)}}^{(+)} + \jump{U}^{(-)} * \av{\sigma^{(1)}}^{(+)} - \av{\Sigma} * \jump{u^{(1)}}^{(+)} - \av{\Sigma} * \jump{u^{(1)}}^{(-)}\nonumber\\&= 
- \jump{U} * P^{(-)} - \av{U} * Q^{(-)} - \av{U} * Q^{(+)}.
\end{align}
Taking the Fourier transform in $x$ yields
\begin{align}
\jump{\overline{U}}^{+}(\xi) \av{\overline{\sigma}^{(1)}}^{+}(\xi) + \jump{\overline{U}}^{-}(\xi) \av{\overline{\sigma}^{(1)}}^{+}(\xi) 
- \av{\overline{\Sigma}}(\xi) \jump{\overline{u}^{(1)}}^{+}(\xi) - \av{\overline{\Sigma}}(\xi) \jump{\overline{u}^{(1)}}^{-}(\xi) \nonumber\\ = 
- \jump{\overline{U}}(\xi) \overline{P}^{-}(\xi) - \av{\overline{U}}(\xi) \overline{Q}^{-}(\xi) - \av{\overline{U}}(\xi) \overline{Q}^{+}(\xi).
\end{align}
We now make use of the transmission conditions
\begin{align}
\jump{\overline{U}}^{-}(\xi)= \kappa \av{\overline{\Sigma}}(\xi), \qquad \jump{\overline{u}^{(1)}}^{+}(\xi) = \kappa \av{\overline{\sigma}^{(1)}}^{+}(\xi) 
- \kappa \overline{P}^{+}(\xi), 
\end{align}
thus obtaining
\begin{align}
\jump{\overline{U}}^{+}(\xi) &\av{\overline{\sigma}^{(1)}}^{+}(\xi) - \av{\overline{\Sigma}}(\xi) \jump{\overline{u}^{(1)}}^{-}(\xi) = 
- \jump{\overline{U}}(\xi) \overline{P}^{-}(\xi) - \av{\overline{U}}(\xi) \overline{Q}^{-}(\xi)- \kappa \av{\overline{\Sigma}}(\xi) \overline{P}^+(\xi) 
- \av{\overline{U}}(\xi) \overline{Q}^{+}(\xi).
\end{align}
The same reasoning used in Section \ref{plane:section:betti}, allows us to derive the integral representation for $\Delta \sigma_0$ in the form
\begin{align}
\Delta \sigma_0 = -\frac{1}{2}\sqrt{\frac{\mu_0}{\pi}} 
&\left\{
\int_{-\infty}^\infty \biggl[ \xi \jump{\overline{U}}(\xi) \overline{P}^-(\xi) + \xi \av{\overline{U}}(\xi) \overline{Q}^-(\xi) \right] \mathrm{d}\xi + \int_{-\infty}^\infty \left[ \kappa \xi \av{\overline{\Sigma}}^-(\xi) \overline{P}^+(\xi) + \xi \av{\overline{U}}(\xi) \overline{Q}^+(\xi) \right] \mathrm{d}\xi
\biggr\}.\label{deltaa0expression}
\end{align}
This important constant has an immediate physical meaning. If $\Delta \sigma_0=0$ then the defect configuration is neutral; its presence causes zero perturbation to the leading order of tractions at the crack tip. Otherwise, if $\Delta \sigma_0<0$, the presence of the defect causes a reduction in the crack tip traction and so shields the crack from propagating further. Finally, if $\Delta \sigma_0>0$ then the defect causes an amplification effect and so can be considered to be encouraging the propagation of the main crack.

\section{Numerical results}\label{section:numerical}

\begin{figure}[t!]
\begin{center}
      \includegraphics[width=0.8\linewidth]{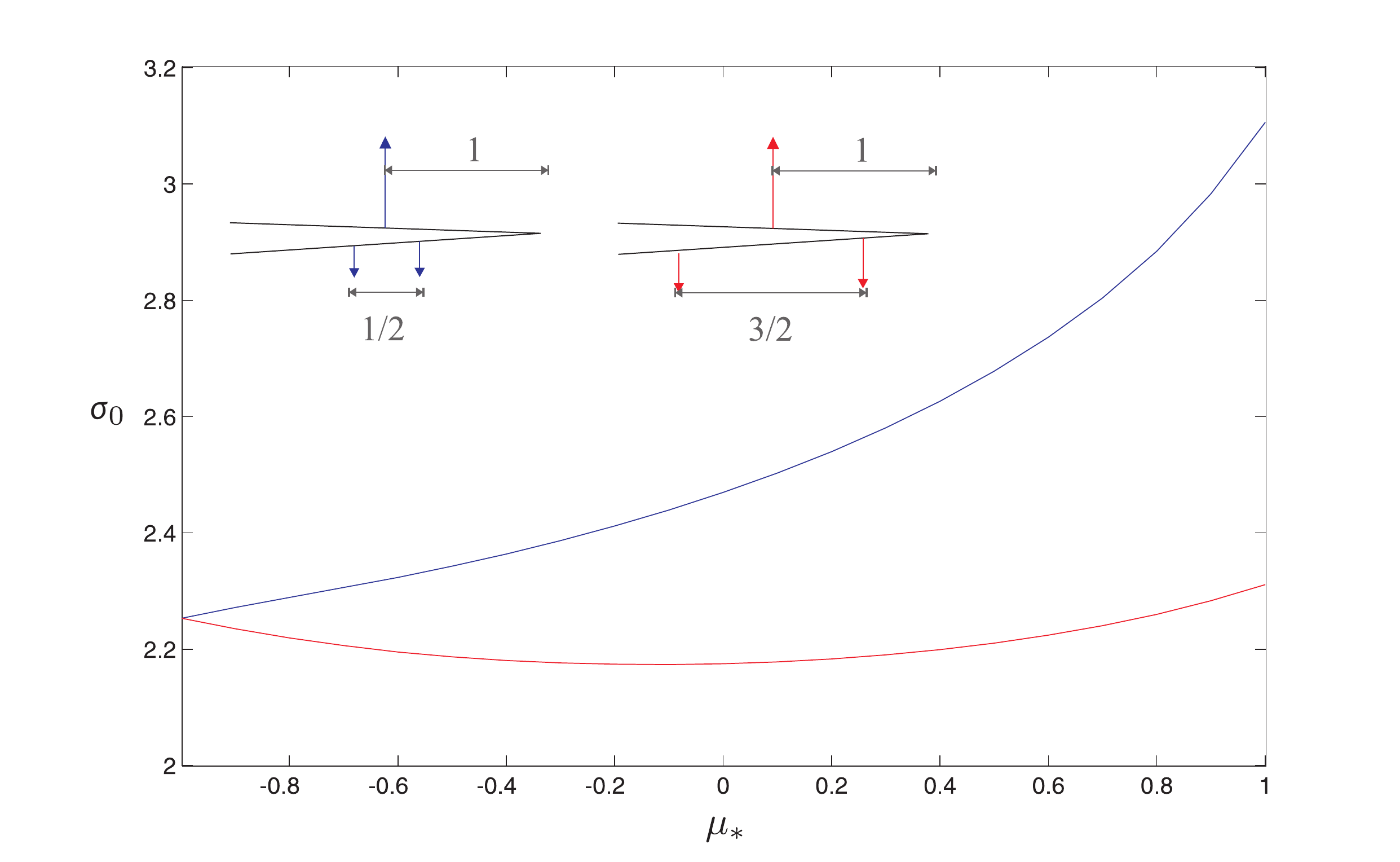}
\end{center}
\vspace{-2em}\caption{Plot of $\sigma_0$ against $\mu_*$. Both cases plotted here use the parameters $\kappa_*=1$ and $a=1$, but with different values for $b$, which controls the separation between the point loadings. The red plot has $b=3/4$ while the blue plot uses $b=1/4$.}
\label{plane:a0vsmustar}
\end{figure}

\subsection{Computations of $\sigma_0$}
In this section we present results of computations obtained by following the methods previously described in this paper. All results have been computed using MATLAB.

Figure \ref{plane:a0vsmustar} plots $\sigma_0$ against $\mu_*$, showing how the constant from the asymptotic expansion at the crack tip $\sigma_0$ varies with differently contrasting stiffnesses of materials. Recalling that 
\begin{equation}
  \mu_*=\frac{\mu_1-\mu_2}{\mu_1+\mu_2},
\end{equation}
we note that when $\mu_*$ is near to $-1$, this corresponds to $\mu_2\gg\mu_1$. That is, the material occupying the region below the crack is far stiffer than the material above the crack. As this limit is approached, the precise locations of the point loadings on the lower face of the crack decrease in importance, since the material becomes sufficiently stiff for the material to act as an almost rigid body; this explains the meeting of the two lines at $\mu_*=-1$.

\begin{figure}
\begin{center}
      \includegraphics[width=0.9\linewidth]{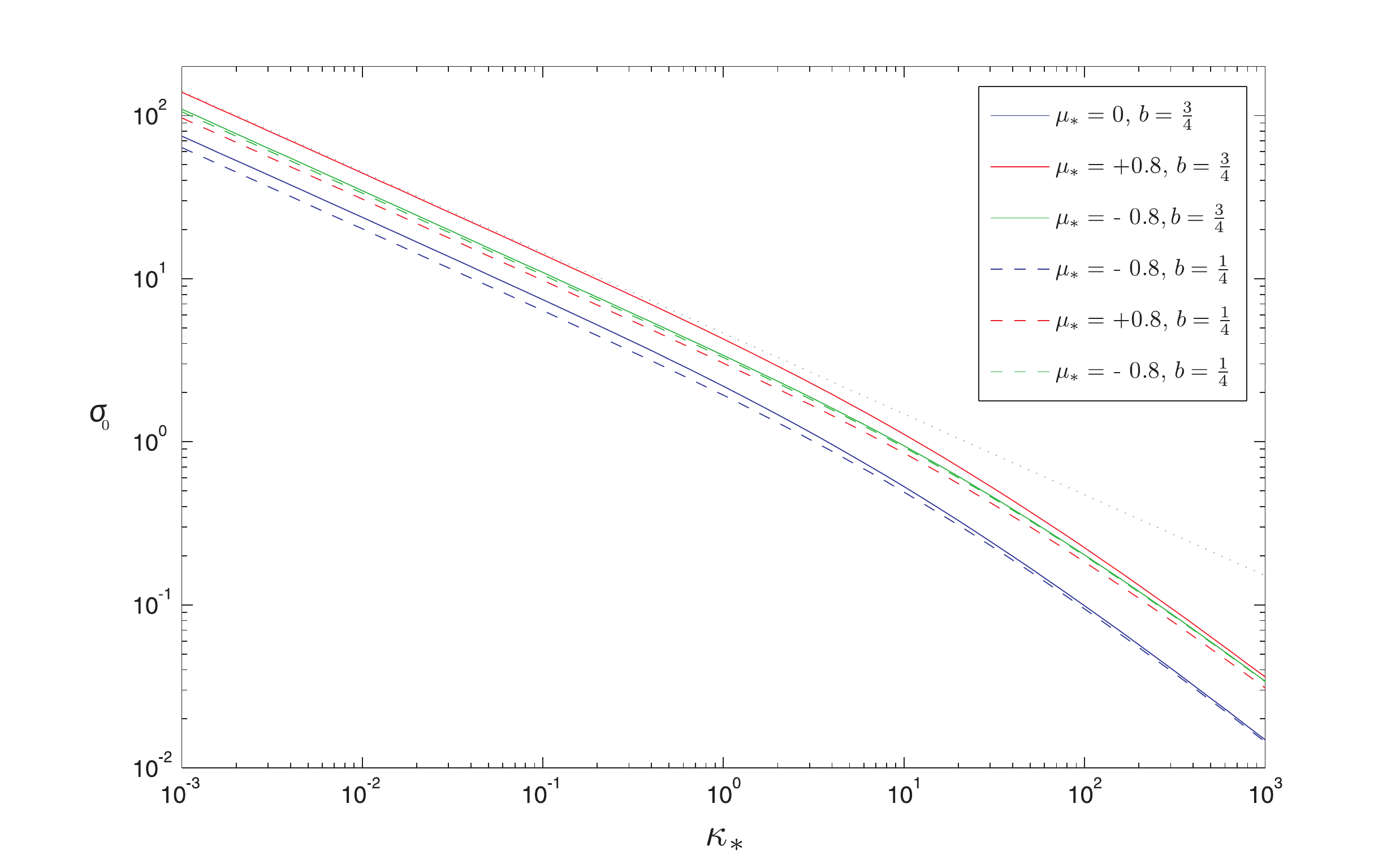}
\end{center}
\vspace{-2em}
\caption{Log-log plot of $\sigma_0$ against $\kappa_*$ for differently contrasting materials.}
\label{plane:tramlines}
\end{figure}

In Figure \ref{plane:tramlines} we present a log-log plot of $\sigma_0$ against $\kappa_*$, the dimensionsless parameter of interface imperfection defined as
$\kappa_*=\kappa(\mu_1+\mu_2)/a$.
This has been computed for different values of $\mu_*$ (describing the contrast in material stiffnesses) and also for different values of $b$ (describing the separation distance between the point loadings) while keeping $a$ fixed ($a=1$). The solid lines correspond to $b=\frac{3}{4}$ while dotted lines represent $b=\frac{1}{4}$ and different colours correspond to different values of $\mu_*$: green corresponds to $\mu_*=-0.8$, blue to $\mu_*=0$ and red to $\mu_*=+0.8$.

Bearing in mind our remarks regarding Figure \ref{plane:a0vsmustar}, we would expect that changing the value of $b$ would have the greatest impact for values of $\mu_*$ near +1. This is indeed the case in Figure \ref{plane:tramlines}.

Also plotted in Figure \ref{plane:tramlines} is a grey dotted line that is tangent to the curves (which run parallel) as $\kappa_*\to0$; this tangent has slope $-\frac{1}{2}$, indicating that $\sigma_0=O(\kappa_*^{-1/2})$ as $\kappa_*\to0$. As $\kappa_*\to0$, the interface becomes almost perfect, and so the square-root behaviour associated with fields near crack tips in the perfect interface setting is not unexpected. Moreover, as $\kappa\to+\infty$, the curves on the log-log plot have slope $-1$, implying that $\sigma_0=O(\kappa_*^{-1})$ as $\kappa_*\to+\infty$. 
%Although the plot only shows values of $\kappa_*$ between $10^{-3}$ and $10^3$, calculations have been performed between $10^{-9}$ and $10^9$ which verify these asymptotic behaviours.

{Computations analogous to those presented in Figure \ref{plane:tramlines} have been performed for smooth asymmetric loadings given by
\begin{equation}\label{plane:nicerloadings}
 p_+^{(-)}(x)=-\frac{4}{9}xe^{2x},\quad p_-^{(-)}(x)=-xe^{3x},\quad x<0;
\end{equation}
we do not present them here since changing the loading to the form (\ref{plane:nicerloadings}) introduces no new features. In the following subsection however, we will detail an approach for comparing $\sigma_0$ against stress intensity factors and will present computations there for both point and smooth loadings.}

\subsection{Comparison of $\sigma_0$ with stress intensity factors from the perfect interface case}
In this subsection we discuss an approach which enables a comparison to be made between imperfect and perfect interface situations.

Comparing the fields directly is not a simple task since in the perfect interface case the stresses become unbounded at the crack tip, exhibiting asymptotic behaviour of $\sigma=O(r^{-1/2})$, $r\to0$. In the imperfect setting, we have derived the leading order of stresses at the crack tip, $\sigma_0$, which is independent of $r$. Moreover, different normalisations may make comparisons difficult.

However, given two particular pairs of materials with {contrast} parameters $(\mu_*)_1$ and $(\mu_*)_2$ say, we might expect the dimensionless ratios of stress intensity factors ${(K_{III}^{(0)})_1}/{(K_{III}^{(0)})_2}$ (from the perfect interface case) and ${(\sigma_0)_1}/{(\sigma_0)_2}$ (imperfect case) to be similar for small $\kappa_*$.
% % \begin{minipage}{0.5\linewidth}
% \begin{figure}
% \begin{center}
%       \begin{minipage}{0.45\linewidth}\includegraphics[width=1.1\linewidth]{ratios_splined.eps}\end{minipage}
%       \begin{minipage}{0.45\linewidth}\includegraphics[width=1.1\linewidth]{smooth_sif_comparison.eps}\end{minipage}
% \end{center}
% \caption{Plot of the ratio $r$ as defined in (\ref{plane:ratior}) for four different values of $(\mu_*)_2$.}
% \label{plane:ratio}
% \end{figure}

\begin{figure}[ht]
\begin{minipage}[b]{0.5\linewidth}
\centering
\includegraphics[width=\textwidth]{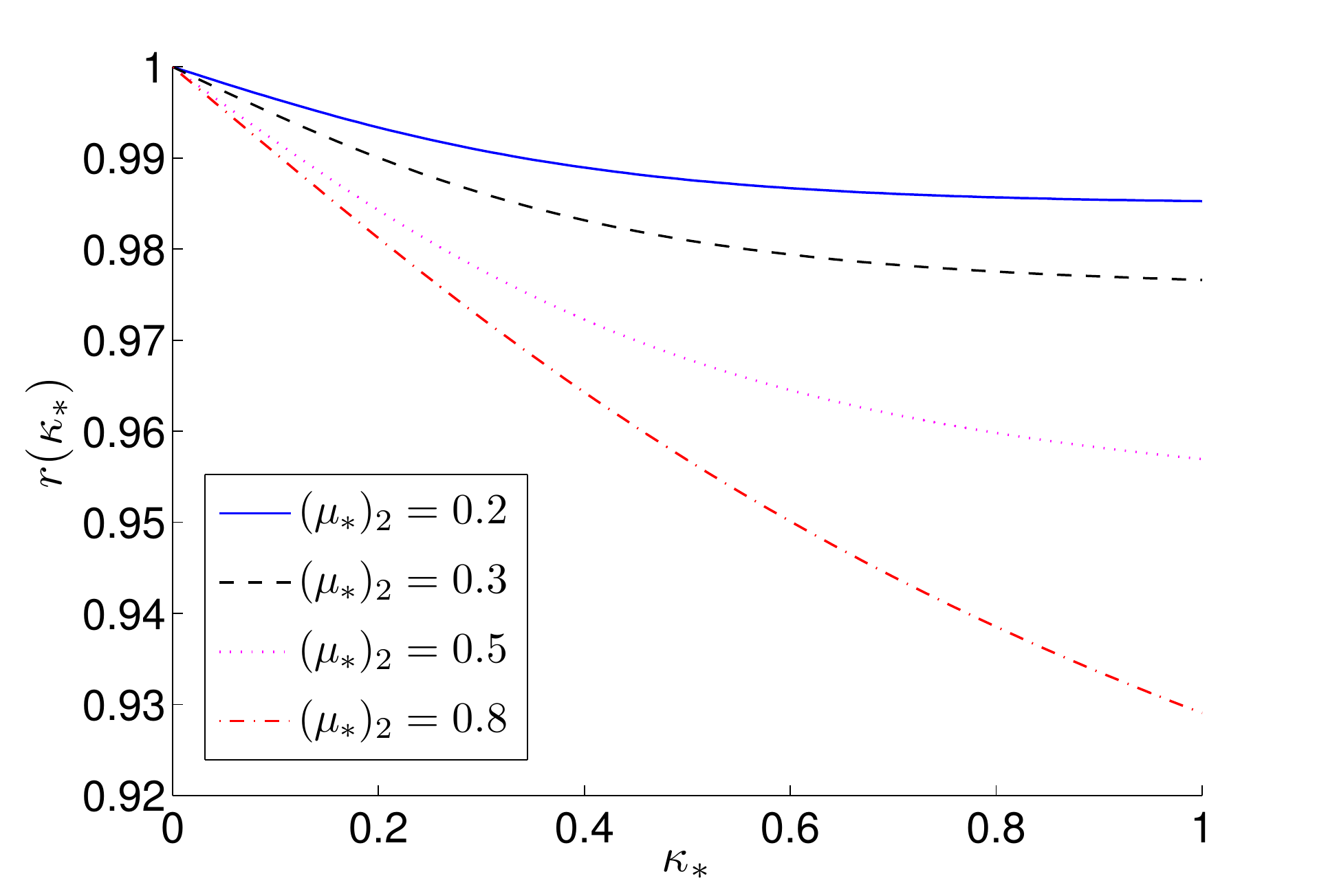}
\caption{{Plot of the ratio $r$ as defined in (\ref{plane:ratior}) for four different values of $(\mu_*)_2$ with point loadings of the form (\ref{pointloadingsdefn}) acting on the crack faces.}}
\label{plane:ratio}
\end{minipage}
\hspace{0.1cm}
\begin{minipage}[b]{0.5\linewidth}
\centering
\includegraphics[width=\textwidth]{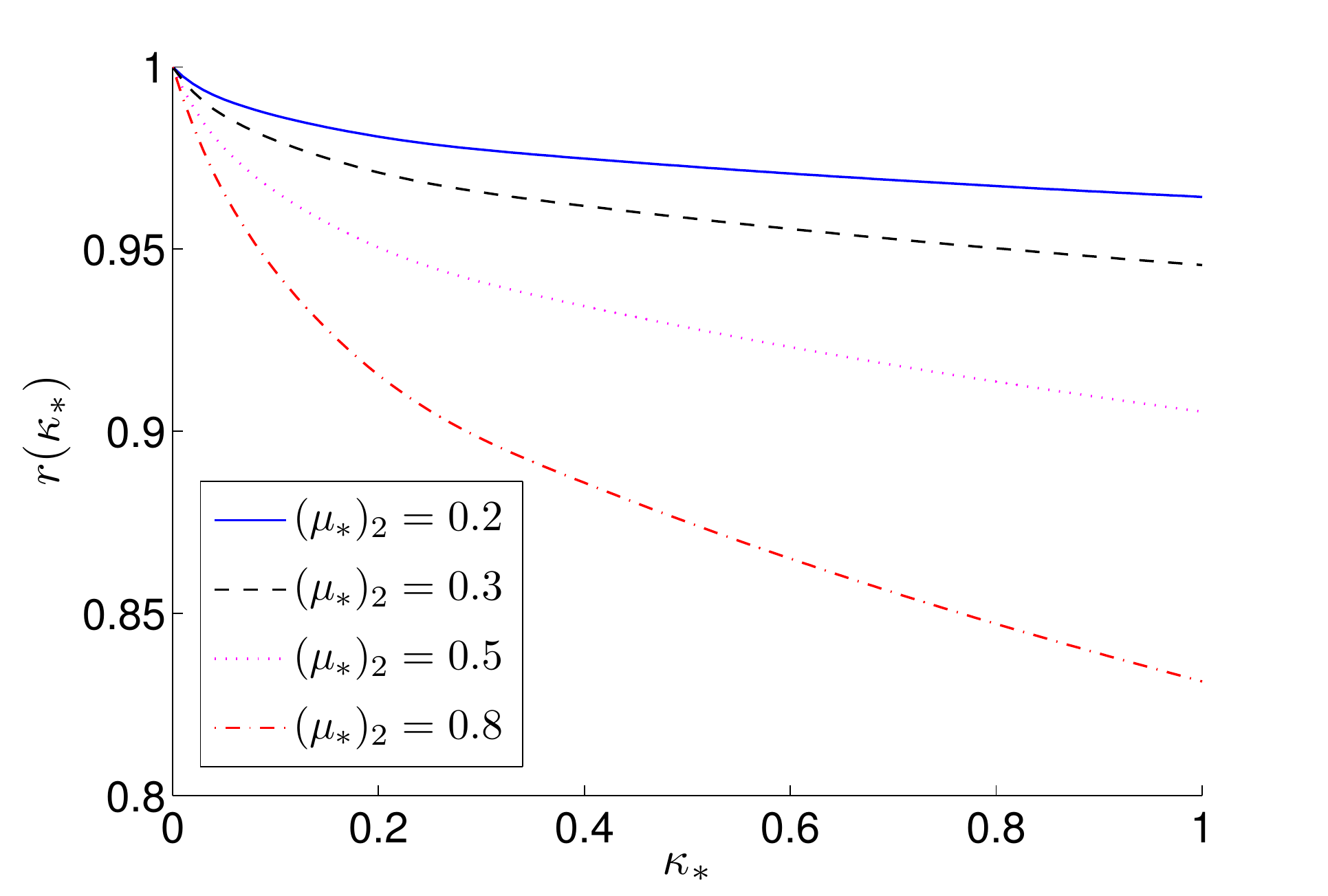}
\caption{{Plot of the ratio $r$ as defined in (\ref{plane:ratior}) for four different values of $(\mu_*)_2$ with smooth loadings of the form (\ref{plane:nicerloadings}) acting on the crack faces.}}
\label{plane:ratiosmooth}
\end{minipage}
\end{figure}

% \end{minipage}
In the perfect interface case, the stress intensity factor (derived in \cite{Piccolroaz2012}) is given by
\begin{equation}
 K_{III}^{(0)}=-\sqrt{\frac{2}{\pi}}\int\limits_0^\infty\left\{\av{p}(-r)+\frac{\mu_0}{2}\jump{p}(-r)\right\}r^{-1/2}\mathrm{d}r.
\end{equation}
As derived earlier in Section \ref{plane:section:betti}, {the leading order of tractions near the crack tip in the imperfect interface case is given by 
\begin{equation}
 \sigma_0^{(0)}=\frac{1}{2}\sqrt{\frac{\mu_0}{\pi}}\int\limits_{-\infty}^\infty\xi\left(\bar{\ljump U\rjump}(\xi)\bar{\langle p\rangle}(\xi)+\bar{\langle U\rangle}(\xi)\bar{\ljump p\rjump}(\xi)\right)\mathrm{d}\xi;
\end{equation}
we emphasise that this quantity depends heavily upon the extent of interface imperfection, characterised by the dimensionless parameter $\kappa_*$.}

Figure \ref{plane:ratio} plots the ratio
\begin{equation}\label{plane:ratior}
 r(\kappa_*)=\frac{(\sigma_0(\kappa_*))_1/(\sigma_0(\kappa_*))_2}{(K_{III}^{(0)})_1/(K_{III}^{(0)})_2}
\end{equation}
for $0<\kappa_*<1$ with $(\mu_*)_1=0$ fixed and for four different values of $(\mu_*)_2$. The loadings used are balanced; a point loading on the upper crack face at $x=-1$ is balanced by two equal loadings at $x=-1.25$ and $x=-0.75$.

We see from the plot that as $\kappa_*\to0$, $r(\kappa_*)\to1$. This provides some verification of the accuracy of our computations {for asymmetric point loadings} and demonstrates that the comparison of ratios approach for small $\kappa$ again the perfect interface case is useful.

%{Computations have also been performed for smooth loadings, specifically

{Figure \ref{plane:ratiosmooth} plots the ratio $r(\kappa_*)$ for the smooth asymmetric loadings described by (\ref{plane:nicerloadings}). We see that $r(\kappa_*)\to1$ as $\kappa_*\to0$, thus demonstrating that $\sigma_0$ is comparable with stress intensity factors for smooth loadings as well as point loadings.}
% Together, Figures \ref{plane:ratio} and \ref{plane:ratiosmooth} suggest that results for point loadings are qualitatively similar to those for smooth loadings.

\subsection{Computation of $\Delta \sigma_0$}
\begin{figure}
\begin{center}
      \includegraphics[width=0.7\linewidth]{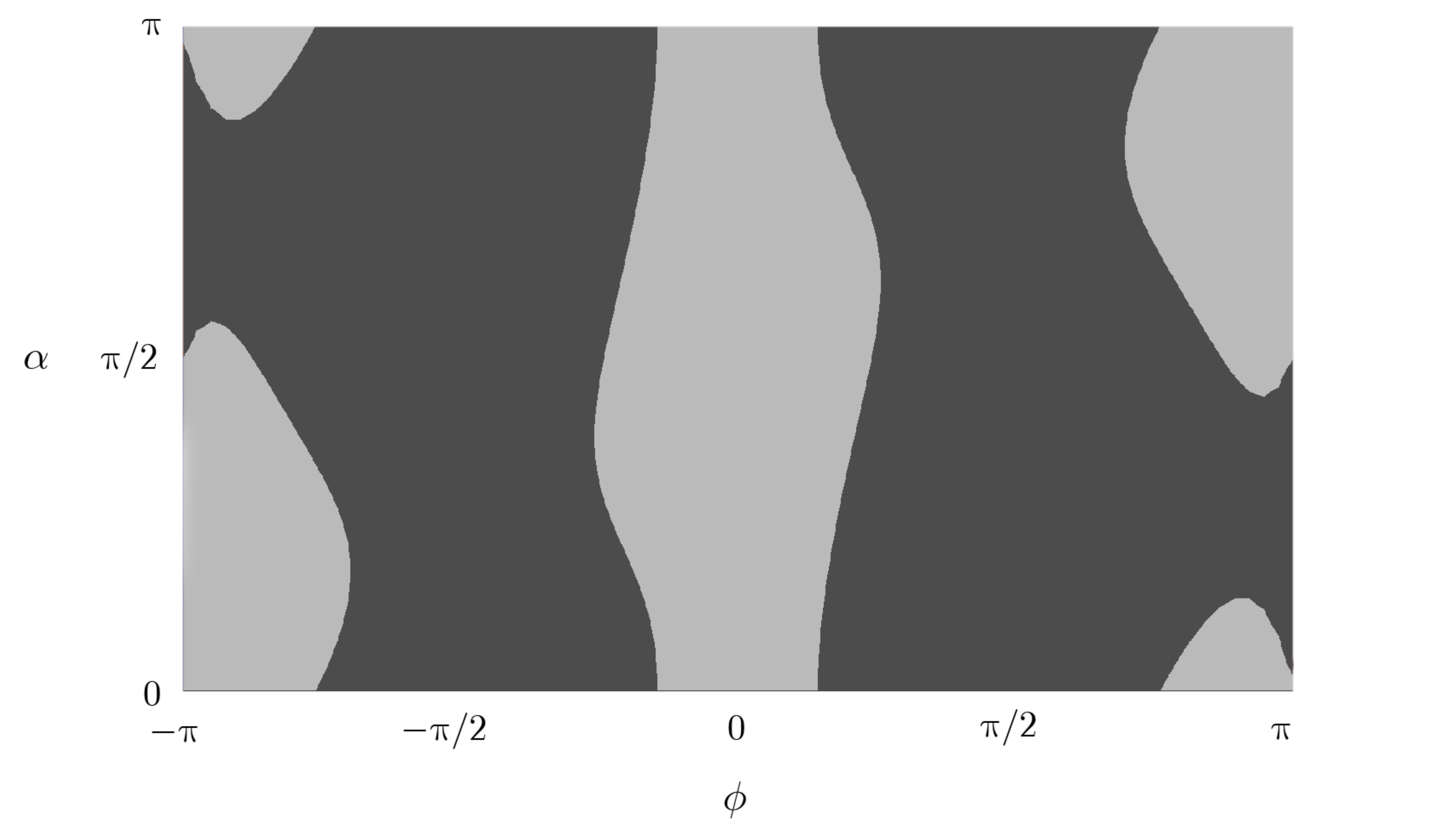}
\end{center}
\vspace{-2em}\caption{Plot of the sign of $\Delta \sigma_0$ for varying $\alpha$ and $\phi$. The darker shaded areas are those $(\phi,\alpha)$ for which $\Delta \sigma_0>0$ while paler regions have $\Delta \sigma_0<0$.}
\label{plane:regions}
\end{figure}

We now present numerical results for the perturbed problem computed using MATLAB. Figure \ref{plane:regions} shows the sign of $\Delta \sigma_0$ for a specific configuration. To reduce the computational task here, we have used smooth loadings with the tractions on the upper and lower crack faces of the form (\ref{plane:nicerloadings}); the imperfect interface has $\kappa_*=1$. {The results presented in Figures \ref{plane:ratio} and \ref{plane:ratiosmooth} demonstrate that results for point loadings and smooth loadings are qualitatively similar. We emphasise however that the perturbation methods described in Section \ref{section:pert} are applicable to both smooth and point loadings.} The inclusion is stiff, with the contrast between the internal and external materials of the inclusion given by ${\nu_*}=5$.
% \begin{figure}
%  \begin{center}
%   \begin{table}
%   \begin{center}
%     \begin{tabular}{|c|c|c|c|c|c|c|c|c|c|}
%         \hline
%         $\mu_1$ & $\mu_2$       & $\mu_*$ & $\mu_{\text{in}}$ & $\ell_a$ & $\ell_b$ & $\kappa$ & $a_1$ & $a_2$ \\ \hline
%         1       & $1$ & $0$ & 10                      & 0.01     & 0.02     & 0.6      & 2     & 3     \\
%         \hline
%     \end{tabular}
%     \end{center}
%     \caption{Table of parameters used for the computation of $\Delta \sigma_0$.}\label{parameterstable}
%   \end{table}
%  \end{center}
%  \caption{Table of parameters used for the computation of $\Delta \sigma_0$.}
% \end{figure}

The figure clearly shows the regions for which crack growth is encouraged or discouraged for this configuration. However, we make the observation that different analysis should be sought when $\phi$ is particularly close to zero since this corresponds to the crack being placed near the imperfect interface which contradicts the assumption made before equation (\ref{boundary_layer_1}).

\section{{Conclusions}}
{The imperfect interface weight function techniques presented here allow for the leading order out-of-plane component of stress and the displacement discontinuity near the crack tip to be quantified. 
The displacement discontinuity can serve as an important parameter in fracture criteria for imperfect interface problems; we demonstrated that, in the limiting case as the extent of imperfection tends towards zero, the criterion is consistent with classical criteria based on the notion of the stress intensity factor.  Perturbation analysis further enables us to correct the solution to account for the presence of a small inclusion. The techniques presented enable us to determine whether the defect's presence shields of amplifies the propagation of the main crack.}

{Although we have presented computations in this paper for the situation where only one such inclusion is present and the inclusion is elliptical, we stress that the technique is readily applicable to geometries containing {any number} of small independent defects, provided a corresponding dipole matrix for each inclusion is used. Indeed, even homogenisation-type problems for composite materials with the main crack lying along a soft imperfect interface of the composite could be tackled using the described techniques. Moreover, similar analysis could be conducted for more general problems, for instance Mode I/Mode II analysis and for various different types of imperfect interface (see for example \cite{Mish1997,MishKuhn2001,Linkov}).
}

\section*{Acknowledgements}
AV and GM respectively acknowledge support from the FP7 IAPP projects PIAP-GA-2009-251475-HYDROFRAC and PIAP-GA-2011-286110-INTERCER2.  AP would like to acknowledge the Italian Ministry of Education,
University and Research (MIUR) for
the grant FIRB 2010 Future in Research ``Structural mechanics models for
renewable energy applications''
(RBFR107AKG).

\end{document}